\documentclass[final,3p,times]{elsarticle}
\usepackage{subcaption}
\usepackage{amsmath, amssymb, amsfonts, amsthm}
\usepackage{graphicx} 
\usepackage{tikz}
\usetikzlibrary{decorations.markings}
\usepackage{pgfplots}
\usepackage{verbatim, soul} 
\usepackage{hyperref}
\usepackage[percent]{overpic}
\usepackage[normalem]{ulem}
\pgfplotsset{compat=1.18}
\makeatletter
\def\th@remark{%
\normalfont 
\thm@headfont{\bfseries} 
\itshape 
}
\makeatother

\newtheorem{theorem}{Theorem}
\newtheorem{proposition}{Proposition}
\newtheorem{lemma}{Lemma}
\newtheorem{corollary}{Corollary}
\theoremstyle{definition}
\newtheorem{definition}{Definition}
\newtheorem{assumption}{Assumption}
\theoremstyle{remark}
\newtheorem{remark}{Remark}

\begin{document}
  \begin{frontmatter}

   \title{
   Symmetry Breaking, Hysteresis, and Convergence to the Mean Voter\\ in two-party Spatial Competition
  }

    \author[1]{Daniel Miranda Machado }
    \ead{daniel.miranda@ufabc.edu.br} \ead[url]{https://danielmiranda.prof.ufabc.edu.br/}

    \author[1]{Roberto Venegeroles\corref{cor1}}
    \ead{roberto.venegeroles@ufabc.edu.br}

    \cortext[cor1]{Corresponding author}

   \affiliation[1]{organization={Center for Mathematics, Computation and Cognition, Universidade Federal do ABC (UFABC)},
    addressline={Avenida dos Estados, 5001},
    city={Santo André},
    postcode={09280-560},
    state={SP},
    country={Brazil}}

    \begin{abstract}
      Classical spatial models of two-party competition typically predict convergence to the median voter, yet real-world party systems often exhibit persistent and asymmetric polarization. We develop a spatial model of two-party competition in which voters evaluate parties through general satisfaction functions, and a width parameter $q$ captures how tolerant they are of ideological distance. This parameter governs the balance between centripetal and centrifugal incentives and acts as the bifurcation parameter governing equilibrium configurations. Under mild regularity assumptions, we characterize Nash equilibria through center-distance coordinates, which separate the endogenous political center from polarization. When the voter density is symmetric, the reduced equilibrium condition exhibits a generic supercritical pitchfork bifurcation at a critical value $q_{c}$. Above $q_{c}$, the unique stable equilibrium features convergence to the center, recovering the classical median voter result, whereas below it two symmetric polarized equilibria arise. Asymmetry in the voter distribution unfolds the pitchfork, producing drift in the endogenous center and asymmetric polarized equilibria. The resulting equilibrium diagram has an S-shaped geometry that generates hysteresis, allowing polarization to persist even after tolerance returns to levels that would support convergence in a symmetric environment. In the high-tolerance regime, we show that the unique non-polarized equilibrium converges to the mean of the voter distribution, while the median is recovered only under symmetry. Hence, unlike the Hotelling--Downs model, where convergence to the median is universal, the median voter appears here as an asymptotic benchmark rather than a robust predictor.
    \end{abstract}

  \end{frontmatter}

  \section{Introduction}

  Spatial models have long provided a unifying framework for analyzing competition
  among economic agents and political actors. Originating in Hotelling’s seminal
  contribution to industrial organization \cite{Hotelling1929}, spatial models describe
  strategic interaction in terms of locations chosen in an abstract characteristic
  space. In his classic 1929 article, Hotelling showed that firms competing in a
  linear market with horizontally differentiated products tend to move toward the
  center in equilibrium, a phenomenon now known as the principle of minimum
  differentiation. This insight demonstrated that the geometry of the underlying
  space can shape competitive outcomes, even when firms share identical cost
  functions and face equivalent competitive incentives.

  The spatial perspective was later extended to political competition by Downs
  \cite{Downs1957}, whose 1957 model introduced the assumption that voters choose
  the party or candidate whose policy position is closest to their own ideal point.
  In this setting, parties maximize vote shares rather than profits, but the
  logic of strategic positioning remains structurally similar to Hotelling’s
  framework. Downs’ formulation gave rise to the classical median voter prediction,
  according to which vote-maximizing parties converge to the electoral median
  under suitable regularity conditions on voter preferences and the policy space.

  Although classical spatial models predict a tendency toward convergence to the median voter, the empirical behavior of political elites often displays substantial
  and persistent deviations from this idealized case. This indicates that
  centripetal incentives emphasized in the classical framework reflect only part of the strategic environment. Several centrifugal forces can sustain policy separation. Party leaders must account for the mobilization and participation of activists
  and core supporters, whose ideal points may lie far from the political center \cite{aldrich1983,baron1993}.
  Candidates also face limits to their ideological mobility, because their choices are constrained by prior commitments and reputational concerns \cite{hinichmunger1994,alesina1988,alesinacukierman1990,alesinarosenthal1995}.
  These considerations imply that parties often face trade-offs between moving
  toward the center and maintaining the support of groups essential to their
  organizational viability.

  The purpose of this paper is to study a spatial model of two-party electoral competition that captures the strategic tension between convergence toward the
  center and divergence away from it. The model is built upon general voter satisfaction functions for the two parties and introduces a key parameter that governs how tolerant voters are of candidates whose ideological positions differ from their own. This parameter plays a central role in shaping the balance between centripetal and centrifugal forces in the strategic landscape. Mathematically, it serves
  as the bifurcation parameter that mediates the parties’ equilibrium positions in the ideological space, allowing the model to generate both proximity to the center and sustained ideological separation as admissible equilibrium outcomes.

  Recent applications of this modeling framework highlight its potential across different
  empirical and theoretical contexts. In \cite{YAKM}, the authors examine a one-dimensional
  dynamical version of the model using a Gaussian specification for voter satisfaction
  and apply it to historical data on U.S. legislators, estimating voter tolerance levels and tracking the evolution of party positions over time. In
  \cite{Venegeroles2025}, the model is extended to a multidimensional ideological
  space with a multivariate Gaussian specification, yielding analytical
  conditions under which parties either converge toward the center or position
  themselves farther from it. In that framework, the effective dimensionality of
  the ideological space emerges as a central theoretical factor governing the
  extent of party divergence, providing a mathematical account of how shifts in ideological
  dimensionality can shape patterns of partisan polarization.

  Here, we develop an abstract analytical framework for one-dimensional ideological space that accommodates a broad class of satisficing kernels and voter densities, rather than fixing a particular functional form. Second, within this framework, we characterize the structural terms that govern the local equilibrium structure as voter tolerance changes. Third, we derive verifiable conditions that
  ensure the occurrence of a pitchfork bifurcation and produce explicit examples with hysteresis. Finally, we analyze the high–tolerance regime ($q \to \infty$)
  and show that the game admits a unique non–polarized equilibrium, which converges to the \emph{mean} of the voter distribution in the asymmetric case and to the \emph{median} (coinciding with the mean) when the voter density is
  symmetric, thereby linking the geometry of equilibrium branches to economically meaningful properties of voter satisfaction and its distribution.

The paper is organized as follows. Section~\ref{sec:model} introduces the satisficing spatial model, the center–distance coordinates, and the standing assumptions under which the analysis is conducted. Section~\ref{sec:main-results} presents the main equilibrium configurations and bifurcation scenarios, distinguishing between symmetric and asymmetric voter distributions and outlining the qualitative structure of the resulting equilibria. Section~\ref{sec:general}
develops the general bifurcation analysis and provides the proofs for the generic case, while Section~\ref{sec:highq-proof} establishes the convergence to the mean in the high–tolerance limit. Section~\ref{sec:asymexamples} illustrates the effects of
asymmetry and hysteresis through a Gaussian mixture example. Section~\ref{sec:symmetric}
specializes the theoretical framework to the symmetric benchmark, deriving explicit integral formulas for the bifurcation coefficients, which are applied to Gaussian and
Cauchy specifications in Section~\ref{sec:symexamples}. Section~\ref{sec:conclusion}
concludes.

  \section{The Model} \label{sec:model}

  We assume that voters are distributed along a left--right spectrum over the real line, which we refer to as the ideological space. We consider a two-party
  system, with parties indexed by $i\in\{1,2\}$. Each party $i$ chooses a position $u_{i}$ within this ideological space, and we use $j\neq i$ to denote its rival. We also write a generic pair of party positions as $(u,v)=(u_{1},u_{2})$
  when convenient. In the context of U.S. elections, the left side of the spectrum is typically associated with liberal orientations, whereas the right side is associated with conservative ones. Hence, the distance of a voter or a party
  from the center of this spectrum captures the degree to which they are liberal or conservative, allowing us to interpret their location as a measure of ideological
  intensity.

  A voter at position $x\in\mathbb{R}$ assigns a probability $s_{i}(x,u_{i})$ of
  being satisfied with party $i$. The following assumption specifies the
  structure we impose on this function:

  \begin{assumption}
    \label{assup1} We assume that voter satisfaction with party $i$ at position $u
    _{i}$ is given by a function $s_{i}(x,u_{i})$ satisfying:

    \begin{enumerate}
      \item[(a)] \textit{Spatial structure}: $s_{i}(x,u_{i})$ depends only on the
        ideological distance $|x-u_{i}|$, is continuous and strictly unimodal with
        a unique global maximum attained at $x=u_{i}$, and satisfies
        $s_{i}(x,u_{i})\to 0$ as $|x-u_{i}|\to\infty$.

      \item[(b)] \textit{Functional form}: there exists a strictly unimodal function
        $f:\mathbb{R}_{+}\to[0,1]$ with global maximum $f(0)=1$ and a width
        parameter $q>0$ such that
        \[
          s_{i}(x,u_{i}) = f_{q}(|x-u_{i}|), \qquad f_{q}(t) = f(t/q).
        \]

      \item[(c)] \textit{Smoothness}: the function $f$ satisfies
        \[
          f \in C^{4}(\mathbb{R}_{+}), \qquad f'(0)=0, \quad f''(0)<0,\quad \lim_{t\to\infty}
          f(t)=0.
        \]

      \item[(d)] \textit{Integrability}: the derivatives of $f$ satisfy
        \[
          f',\, f'',\, f^{(3)},\, f^{(4)}\,\in L^{1}(0,\infty), \qquad f' \in L^{2}(0,\infty).
        \]
    \end{enumerate}
  \end{assumption}

  \begin{remark}
    \label{rem1} The width parameter $q$ plays two complementary roles in the model.
    At the behavioral level, it measures how tolerant voters are of parties located
    far from their own ideological positions: larger values of $q$ correspond to
    broader partisan appeal, or greater ``inclusiveness'' of a party across the ideological
    space. From a dynamical perspective, we will later see that $q$ also functions
    as a bifurcation parameter, governing qualitative changes in the strategic
    configuration of party positions.
  \end{remark}

  \begin{remark}
    \label{rem2} Our analysis focuses on party competition in which both parties
    operate on equal footing, which motivates the assumption that voters' satisfaction
    with each party can be represented by the same function $f$ and a common width
    parameter $q$, differing only in the ideological positions $u_{i}$ chosen by
    the parties. Nonetheless, the model readily accommodates heterogeneous
    specifications, allowing for party-specific functions $f_{i}$ and width parameters
    $q_{i}$ without altering its general structure.
  \end{remark}

  \begin{remark}
    \label{rem3} The model can also be extended to electoral competition in multidimensional
    ideological spaces, where parameter matrices combine to play a central role
    in a bifurcation analysis that may shed light on the phenomenon of
    increasing party polarization; see~~\cite{Venegeroles2025}.
  \end{remark}

  When two candidates compete, the probability that a voter at $x$ votes for party
  $i$ against the rival $j$ follows the partitioning rule (e.g., \cite{YAKM}):
  \begin{equation}
    \label{eq:pi}p_{i}(x\mid u_{i},u_{j}) = s_{i}(x,u_{i})\bigl[1-s_{j}(x,u_{j})\bigr
    ] + \tfrac12\, s_{i}(x,u_{i})s_{j}(x,u_{j}),
  \end{equation}
  where the first term on the right-hand side represents the expected share of
  voters at position $x$ who are satisfied only with party $i$, and the second term
  represents the equal split of those who are satisfied with both parties. In
  view of Assumption \ref{assup1}(b), satisfaction with each party is evaluated
  independently: since $s_{i}(x,u_{i})$ and $s_{j}(x,u_{j})$ depend solely on each
  party's proximity to the voter, the events of being satisfied with one party
  and with the other have no causal influence on each other. Hence the joint probability
  of being satisfied with $i$ but not with $j$ factorizes into $s_{i}(1-s_{j})$.
  Evidently, there is also a probability $(1-s_{i})(1-s_{j})$ that some voters are
  not satisfied with either party, and therefore abstain from voting.

  Let $\rho(x)$ denote the density of voters along the ideological space
  $\mathbb{R}$. We impose the following regularity conditions on this density:

  \begin{assumption}
    \label{assup2} We assume that $\rho\in L^{1}(\mathbb{R})\cap L^{\infty}(\mathbb{R}
    )$, $\ \int_{\mathbb{R}}\rho(x)\,dx=1$, and that $\rho$ is continuous.
  \end{assumption}

  The expected vote share for party $i$ is therefore given by
  \begin{equation}
    \label{eq:Vi}V_{i}(u_{i},u_{j})=\int_{\mathbb{R}}p_{i}(x\mid u_{i},u_{j})\,\rho
    (x)\,d x.
  \end{equation}
  As satisfaction depends on the width parameter $q$, we make this dependence explicit
  and write the objective functionals $V_{i}$ as the symmetric pair (using
  $(u,v)$ for the positions of parties 1 and 2):
  \begin{equation}
    \label{eq:Vq-both}
    \begin{aligned}
      V_{1,q}(u,v) & =\int_{\mathbb{R}}\!\Bigl[ f_{q}(|x-u|)-\tfrac12 f_{q}(|x-u|)f_{q}(|x-v|)\Bigr]\,\rho(x)\,dx, \\
      V_{2,q}(u,v) & =V_{1,q}(v,u),
    \end{aligned}
  \end{equation}
  reflecting identical satisficing kernels and label symmetry. The following assumption
  establishes the strategic nature of party competition:

  \begin{assumption}
    \label{assup3} Each party $i$ strategically chooses its ideological position
    $u_{i}\in \mathbb{R}$ to maximize its expected vote share $V_{i,q}(u_{i},u_{j}
    )$. Formally,
    \begin{equation}
      u_{i}\in \arg\max_{u \in \mathbb{R}}V_{i,q}(u,u_{j}).
    \end{equation}
  \end{assumption}

\begin{remark}
  \label{rem4}
  The conditions imposed on the voter density $\rho$ in Assumption~\ref{assup2}
  ensure that expected vote shares are mathematically well defined and that the
  maximization problems in Assumption~\ref{assup3} are well posed. Indeed, for
  all $(u_i,u_j)\in\mathbb{R}^2$ the function
  $x \mapsto p_i(x\mid u_i,u_j)\,\rho(x)$ is integrable, so $V_{i,q}(u_i,u_j)$
  is finite and continuous in $u_i$, and dominated convergence applies. Moreover,
  $V_{i,q}(u,u_j)\to 0$ as $|u|\to\infty$, so for each fixed $u_j$ the map
  $u\mapsto V_{i,q}(u,u_j)$ attains a maximum on $\mathbb{R}$, as required in
  Assumption~\ref{assup3}.
\end{remark}

The strategic nature of this electoral game is closely tied to how the parameter $q$ shapes the trade-off between ideological proximity and satisficing. When $q$ is large, the game admits a unique locally stable Nash equilibrium at an endogenous political center. When $q$ falls below the critical threshold $q_{c}$, however, a supercritical pitchfork bifurcation generates two stable off-diagonal equilibria. Variations in $q$ therefore turn a single consensus-seeking configuration into a coordination problem between competing polarized positions.

  \section{Main Results}
  \label{sec:main-results}

  Throughout this section we assume that the satisficing kernel $f$ and the voter
  density $\rho$ satisfy Assumptions~\ref{assup1}--\ref{assup3}. The detailed statements
  and proofs appear in Sections~\ref{sec:general} and~\ref{sec:symmetric}; here
  we summarize the equilibrium structure.

\begin{figure}[h]
    \centering
    \definecolor{plotblue}{RGB}{0, 0, 128} 

    \begin{tikzpicture}[scale=1., >=stealth]

      \def\peakX{-0.5} 
      \def\maxH{1.6} 
      \filldraw[
        draw=black!90!black,
        thick,
        smooth,
        fill=gray!0,
        fill opacity=0.5
      ]
        plot[domain=-3.5:4, samples=200]
        (\x, {\maxH * exp(-(\x < \peakX ? 0.5 : 0.15) * (\x - \peakX)^2)});
      \node[plotblue!0!black, above]
        at
        (-0.5, \maxH+0.2)
        {Voter Density $\rho(x)$};
      \draw[<->, thick, black!100]
        (-3.5,0) -- (4,0)
        node[below=5pt] {};

      \coordinate (Utwo) at (-2, 0); 
      \coordinate (M) at ( 0, 0); 
      \coordinate (Uone) at ( 2, 0); 

      \draw[thick] (Utwo) -- ++(0, 0.15);
      \filldraw (Utwo) circle (1.5pt);
      \node[below=8pt] at (Utwo) {$u_{2}$};
      \node[below=16pt] at (Utwo) {(Party 2)};

      \filldraw[black] (M) circle (1.2pt);
      \node[below=8pt] at (M) {$m$};
       \node[below=16pt] at (M) { (Center)};

      \draw[thick] (Uone) -- ++(0, 0.15);
      \filldraw (Uone) circle (1.5pt);
      \node[below=8pt] at (Uone) {$u_{1}$ };
            \node[below=16pt] at (Uone) {(Party 1)};

      \draw[
        decorate,
        decoration={brace,amplitude=5.7pt, ,raise=7pt},
        thick,
        black!70
      ] (Utwo) -- (M) node[midway,above=11pt] {$d$};

      \draw[
        decorate,
        decoration={brace,,amplitude=5.7pt,raise=7pt},
        thick,
        black!70
      ] (M) -- (Uone) node[midway, above=11pt] {$d$};

      \draw[decorate, decoration={brace,amplitude=8pt,mirror,raise=35pt}, thick]
        (Utwo) -- (Uone)
        node[midway, below=45pt] {Polarization ($2d = u_{1}- u_{2}$)};
    \end{tikzpicture}
    \caption{\textit{Geometric interpretation of the center-distance coordinates.} The diagram illustrates the change of variables defined in Eq.~\eqref{eq:coords}, in which the party positions $u_1$ and $u_2$ are expressed in terms of the system center $m = (u_1 + u_2)/2$ and the half--distance $d = (u_1 - u_2)/2$. The parameter $d$ provides a natural measure of polarization in the bifurcation analysis.}
    \label{fig:center_distance_coords}
  \end{figure}
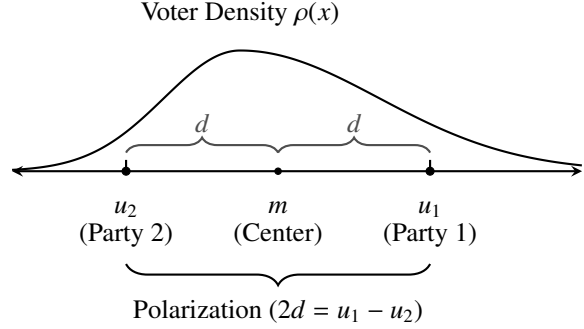

  To analyze the equilibrium configurations of the two-party game, we examine the
  first-order conditions associated with vote-share maximization. Writing the
  original party positions as $u_{1}, u_{2} \in \mathbb{R}$ (or equivalently $(u,
  v)$), it is convenient to introduce the change of variables
  \begin{equation}
    \label{eq:coords}m := \frac{u_{1}+u_{2}}{2}, \qquad d := \frac{u_{1}-u_{2}}{2}
    .
  \end{equation}
  This transformation decouples the strategic interaction into two geometric components:
  the ideological midpoint of the party system ($m$) and its degree of
  polarization ($2d$). In this representation, $m$ is the ideological center of
  gravity of the party system, while $2d$ is the ideological distance between the
  parties, see Fig.~\ref{fig:center_distance_coords}. Hence, the decomposition separates
  the problem of \emph{where} competition takes place ($m$) from \emph{how
  intense} the differentiation is ($d$).

  The parameter $q$ indexes voter tolerance. As $q$ decreases from high values, the equilibrium set undergoes a transition from co-location at the endogenous center to polarization once it falls below a critical level $q_c$. We analyze this transition by decomposing the party positions into the midpoint $m$ and a polarization half--distance $d$, as defined in Eq. \eqref{eq:coords}.

  \subsection{Equilibria in the Symmetric Case}
  We first consider the benchmark case in which the voter density $\rho$ is symmetric around the origin. This symmetry pins the political center at $m=0$ and reduces the equilibrium conditions to a one-dimensional problem for the
  polarization distance $d$: the two-dimensional first--order system collapses
  to a scalar equation of the form
  \[
    \Psi_{q}(d) = 0,
  \]
  where $\Psi_{q}$ is an odd function of $d$ and $q$ is viewed as a parameter. In this setting, and under Assumptions~\ref{assup1}--\ref{assup3}, the reduced
  equation $\Psi_{q}(d)=0$ undergoes a supercritical pitchfork bifurcation at
  $q=q_{c}$, which organizes the generic equilibrium patterns as follows:
  \begin{itemize}
    \item \textit{Convergence regime ($q>q_{c}$).} In a neighborhood of the
      origin, the unique locally stable equilibrium is the centrist configuration
      \[
        (m,d) = (0,0),
      \]
      corresponding to both parties locating at the same point $u_{1}=u_{2}=0$.
      In this regime the model recovers the classical Median Voter Theorem.

    \item \textit{Divergence regime ($q<q_{c}$).} As $q$ decreases past
      $q_{c}$, the centrist equilibrium loses local stability and the system bifurcates
      into two stable, symmetric polarized equilibria; that is,
      \[
        (m,d) = \bigl(0,\pm d^{\ast}(q)\bigr), \qquad d^{\ast}(q)>0,
      \]
      corresponding to the party positions $u_{1} = -u_{2} = d^{\ast}(q)$.

    \item \textit{Bifurcation point $q_{c}$.} The critical threshold
      $q_{c}$ is characterized as the root of the integral equation
      \[
        \int_{0}^{\infty}\left[ f''(t)\Bigl(1-\tfrac12 f(t)\Bigr) + \tfrac12 \bigl
        (f'(t)\bigr)^{2} \right]\rho(q_{c}t)\,dt = 0.
      \]
      Additional transversality and non-degeneracy conditions at $q_{c}$ will
      be addressed in Section \ref{sec:symmetric}; however, their violation is non-generic,
      as it would require the simultaneous nonvanishing of two other independent
      integral equations at the same $q_{c}$.
  \end{itemize}

  These results highlight how, under symmetry, the satisficing framework departs
  from the canonical Hotelling--Downs model. When tolerance is high, parties behave as in the classical theory and cluster at the origin of the ideological space. When tolerance is low, however, locally stable polarization emerges endogenously, despite the absence of any structural asymmetry in the
  electorate. Thus, the Median Voter Theorem is recovered not as a universal prediction, but as a regime-specific outcome that depends on voter tolerance.

  \subsection{Equilibria in the Asymmetric Case}
  Real-world voter distributions need not be symmetric, so we next extend the
  analysis to general densities $\rho$. In this setting, asymmetry acts as a
  geometric imperfection that unfolds the symmetric pitchfork bifurcation, producing
  a qualitatively different local equilibrium structure.

\begin{figure}[t]
\centering
\begin{overpic}[width=9cm,
]{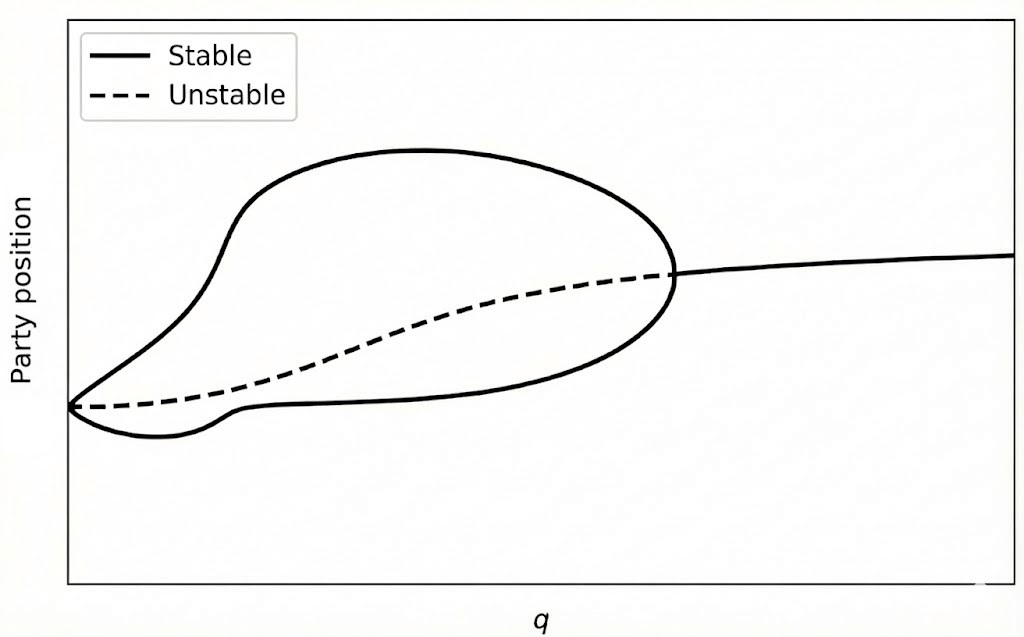}
  \put(60,10){\color{black}\small Asymmetric pitchfork}
\put(70,15){%
  \begin{tikzpicture}
    \draw[dashed,->,line width=0.8pt] (0,0) -- (-0.6,1.2);
  \end{tikzpicture}%
}
  \put(46,18){\color{black}\small $q^{-}$}
  \put(46,22){\color{black}\vector(0,1){10}}
  \put(89,52){\color{black}\small $q^{+}$}
  \put(90,50){\color{black}\vector(0,-1){10}}
    \put(67.2,52){\color{black}\small $q_{c}$}
  \put(67.4,50){\color{black}\vector(0,-1){11}}

   \put(60.2,30){%
    \begin{tikzpicture}
      \draw[dotted, line width=0.9pt] (0,0) circle [radius=0.49cm];
      \fill (0,0) circle [radius=2pt]; 
    \end{tikzpicture}%
  }

\end{overpic}
\caption{Bifurcation diagram illustrating the asymmetric unfolding of the pitchfork bifurcation and the resulting drift of the endogenous center. The values $q^{-}$ and $q^{+}$ delimit the parameter range in which a near-center equilibrium and polarized equilibria coexist.}
      \label{asypitchfork}
\end{figure}
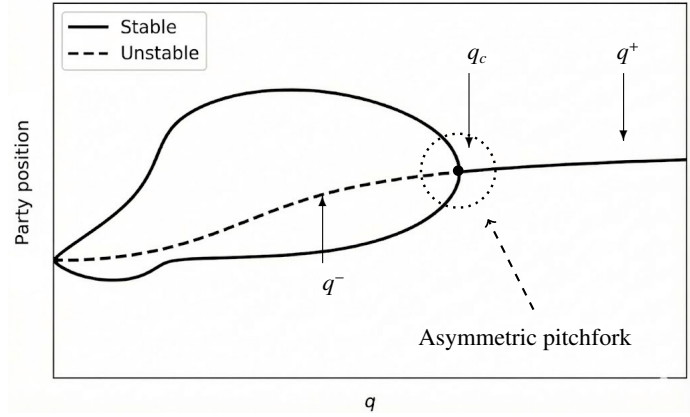

  Let $(f,\rho)$ satisfy Assumptions~\ref{assup1}--\ref{assup3} and assume that
  $\rho$ is not symmetric. Then, generically, there exist critical tolerance levels $0<q^{-}<q^{+}$, as illustrated in Fig.~\ref{asypitchfork}, such that the equilibrium set of the two-party game in a neighborhood of $[q^{-},q^{+}]$ has the following properties:
  \begin{itemize}
    \item \textit{Center drift.} There is no fixed exogenous ``center'': we interpret
      $m_{0}(q)$ as an \textit{endogenous political center} selected by
      equilibrium. In the symmetric case it coincides with the unique median voter,
      but in general it need not equal the median of $\rho$. The midpoint $m_{0}(
      q)$ is endogenously determined and varies smoothly with $q$; in particular,
      for $q$ close to $q^{+}$ there is a unique locally stable equilibrium with
      $(m,d)$ close to $(m_{0}(q),0)$.

    \item \textit{Local asymmetric pitchfork bifurcation.} For $q > q^{+}$ there
      is a unique locally stable equilibrium near $(m_{0}(q),0)$. As $q$ decreases
      past $q^{+}$, two additional equilibria appear, so that for $q^{-}< q < q^{+}$ there are three nearby branches: two locally stable polarized equilibria and one intermediate unstable equilibrium. The transition from the near–center branch to the polarized branches occurs through a local pitchfork–type bifurcation in which both parties move away from the endogenous center $m_{0}(q)$, but to asymmetric positions $(u_{1}(q),u_{2}(q))$ with
      \[
        u_{1}(q)-m_{0}(q) \neq -\bigl(u_{2}(q)-m_{0}(q)\bigr).
      \]
  \end{itemize}

  This result therefore implies that, in asymmetric voter distributions, both parties polarize simultaneously at the same critical point in $q$, but the resulting
  equilibrium is itself asymmetric. One party remains closer to the endogenous center, while the other is pushed into a more extreme niche. In this way, structural
  features of the voter distribution and the level of tolerance interact to determine which party becomes the more extreme pole and how persistent this asymmetric polarization
  is.

  \subsection{High-tolerance Limit: Convergence to the Mean}

 Under Assumptions~\ref{assup1}--\ref{assup3}, together with the genericity hypotheses used in the bifurcation analysis and the additional regularity condition
 $f'' \in L^{1}(0,\infty)\cap L^{\infty}(0,\infty)$, there exists $Q>0$ such that, for every $q \ge Q$, the two-party game admits a unique Nash equilibrium, which is non-polarized: both parties choose the same platform, say
  \[
    (u_{1}(q),u_{2}(q)) \;=\; \bigl(m_{0}(q),m_{0}(q)\bigr).
  \]
  The equilibrium position $m_{0}(q)$ is characterized as the unique centrist solution
  of the first-order condition
  \[
    F_{q}^{+}\bigl(m_{0}(q),0\bigr) \;=\; 0,
  \]
  and depends smoothly on $q$.

  As tolerance increases, the centrist equilibrium $m_{0}(q)$ converges to a simple distributional notion of ``center.'' When the voter density $\rho$ is asymmetric, the high-$q$ limit of $m_{0}(q)$ is given by the \emph{mean} of $\rho$. Under symmetry, however, the equilibrium center is fixed at the origin for all $q>q_{c}$, so the limiting statement becomes trivial: the mean, median, and equilibrium midpoint coincide. Thus, the classical median--voter prediction is recovered only under symmetry and for sufficiently large tolerance. In the asymmetric case, by contrast, the median is generically not an equilibrium position, and the satisficing framework selects the mean—not the median—as the relevant attractor in the high-$q$ limit. 
  
  This stands in contrast with the canonical Hotelling--Downs model, in which parties are forced to converge exactly to the median voter, regardless of the distribution's shape or the degree of voter tolerance. In the satisficing framework, by comparison, convergence to the median is not structurally guaranteed: it arises only under symmetry and sufficiently high tolerance, and it fails whenever tolerance falls below the critical threshold $q_c$. Hence, the median voter serves here as an \emph{asymptotic benchmark} for extremely tolerant voters (or inclusive parties), rather than a robust predictor of equilibrium

  \subsection{Hysteresis}
\label{subsec:hysteresis}

While the symmetric pitchfork bifurcation guarantees a reversible transition between convergence and polarization, the asymmetric unfolding changes the picture in a qualitative way. In the $(q,d)$ plane, the equilibrium set may acquire an S-shaped structure consisting of two stable branches separated by an unstable one. It is precisely this S-shaped geometry that underlies hysteresis.

A convenient way to make the mechanism explicit is to view the evolution of the coordinate $u(t)$ as gradient-like for each fixed tolerance level $q$:
\begin{equation}\label{eq:grad_dynamics}
\dot u = -M(u;q,d)\,\partial_u V(u;q,d),
\end{equation}
where $M(u;q,d)>0$ is an arbitrary mobility function that rescales time but does not affect the location or stability of equilibria. For each fixed $(q,d)$, equilibria are precisely the critical points of the potential,
\[
\partial_u V(u;q,d)=0,
\]
where $d$ is implicitly determined by the equilibrium constraint, so the dynamics are effectively one-dimensional, and local stability corresponds to local minimality (equivalently, $\partial_{uu}V(u;q,d)>0$). In this sense, the ``selection rule'' is built into the dynamics: trajectories are attracted to stable critical points and repelled from unstable ones.

We say the model exhibits hysteresis with respect to tolerance $q$ if the long-run response $u(t)$ is not determined solely by the instantaneous value of $q(t)$, but depends on the history of $q$. Equivalently, there exists a cyclic path of tolerance levels $q(t)$ such that the curve
\[
\{(q(t),u(t)):\ t\in[0,T]\}
\]
forms a nontrivial closed loop in the $(q,u)$ plane.

In the present model, as $q$ decreases, trajectories initially track the stable centrist equilibrium. At a critical value, the centrist equilibrium loses stability, yet the system exhibits a \emph{bifurcation delay}: the trajectory transiently tracks the now-unstable centrist branch due to bifurcation delay, which forms the middle segment of the S-shaped equilibrium curve. This transient tracking reflects a classical bifurcation delay under slow variation of $q$, rather than genuine stability, and should be interpreted as a dynamical effect rather than an equilibrium outcome. When this unstable branch merges with a stable non-centrist branch, the trajectory is forced to leave the centrist curve and lands on a polarized equilibrium.

If tolerance is later restored, polarization does not immediately disappear. Instead, the system remains locked onto the polarized branch until the reverse bifurcation is crossed.

From a political standpoint, this hysteresis mechanism captures a form of political inertia. Once polarization becomes established, simply reverting tolerance to levels that previously ensured convergence may no longer be sufficient to restore it. Past asymmetries can therefore leave persistent imprints on political outcomes, even after the underlying social conditions improve. In contrast to the symmetric case, where transitions are smooth and reversible, the asymmetric unfolding produces history-dependent transitions, making polarization easier to trigger than to undo.

  \
  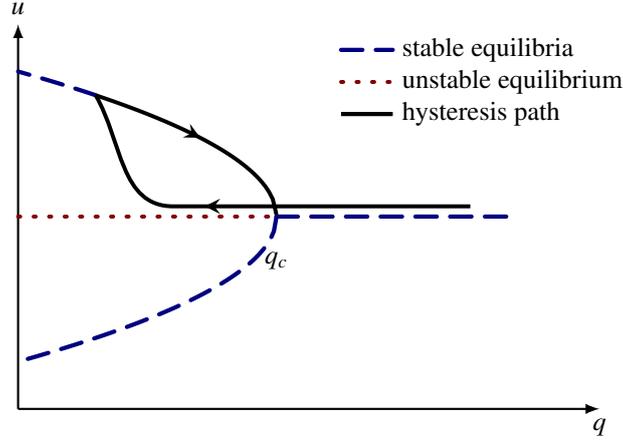
\begin{figure}[h]
    \centering
    \definecolor{plotgreen}{RGB}{0,0,0}
    \definecolor{plotred}{RGB}{129,0,0} 
    \definecolor{plotblue2}{RGB}{0, 0, 128} 

    \tikzset{ midarrow/.style={ postaction={decorate}, decoration={ markings, mark=at position #1 with {\arrow{stealth}} } } }

    \begin{tikzpicture}[>=latex, scale=1.7]
  \draw[->, thick] (0,-1.5) -- (0,1.5) node[above] {$u$};
  \draw[->, thick] (0,-1.5) -- (4.5,-1.5) node[below] {$q$};

  \def\xc{2}
  \node[below] at (\xc, -0.2) {$q_{c}$};

  \tikzset{
    stable/.style={
      line width=1.6pt,
      plotblue2,
      line cap=round,
      dash pattern=on 8pt off 5pt
    },
    unstable/.style={
      line width=1.5pt,
      plotred,
      line cap=round,
      dash pattern=on 0.8pt off 5pt
    }
  }

  \draw[unstable] (0,0) -- (\xc,0);

  \draw[stable] (\xc,0) -- (3.8,0);

  \draw[stable, domain=3.4:4.0, samples=100]
    plot (-\x+2*\xc, {0.8*sqrt(\x-\xc)});
  \draw[stable, domain=\xc:4.0, samples=100]
    plot (-\x+2*\xc, {-0.8*sqrt(\x-\xc)});


  \draw[line width=1.5pt, plotgreen, midarrow=0.6]
    (3.5, 0.08) --
    (1.2, 0.08) to[out=180, in=-60]
    (0.6, {0.8*sqrt(3.4-\xc)});

  \draw[line width=1.5pt, plotgreen, midarrow=0.5]
    plot[domain=3.4:\xc, samples=50]
    (-\x+2*\xc, {0.8*sqrt(\x-\xc)});

  \draw[stable] (2.5,1.3) -- ++(0.4,0);
  \node[right] at (2.9,1.3) {stable equilibria};

  \draw[unstable] (2.5,1.05) -- ++(0.4,0);
  \node[right] at (2.9,1.05) {unstable equilibrium};

  \draw[ultra thick, plotgreen] (2.5,0.8) -- ++(0.4,0);
  \node[right] at (2.9,0.8) {hysteresis path};
\end{tikzpicture}

    \caption{\textit{Asymmetric  supercritical pitchfork and hysteresis:} in the supercritical regime, a fold of large--amplitude polarized equilibrium creates a hysteresis
    loop. The dashed red branch represents unstable equilibria, the navy branches correspond to locally stable equilibria, and the green arrows indicate the
    hysteresis trajectory followed under slow changes in tolerance. Note: The vertical axis represents the equilibrium position $u$ (which proxies for
    polarization).}
    \label{fig:bifurcation_types}
  \end{figure}

  \subsection{Explicit Examples}

 We rigorously characterize the effective linear and cubic coefficients that govern the local bifurcation structure as the tolerance parameter $q$ varies. In particular, we compute the critical threshold $q_{c}$ and determine the nature of the bifurcation in several explicit examples; the corresponding derivations are presented in Sections~\ref{sec:asymexamples} and~\ref{sec:symexamples}. Under these assumptions, the following results hold:
  \begin{itemize}
    \item \textit{Gaussian kernel, Gaussian density.} When both the satisficing
      kernel and the voter density are Gaussian---specifically, $f(x)=e^{-x^{2}/2}$
      and $\rho(x) \propto e^{-x^{2}/2}$---the reduced equation exhibits a supercritical
      pitchfork bifurcation at $q_{c}\approx 0.8074$. For $q < q_{c}$, two polarized
      equilibria emerge continuously from the symmetric one, so that thin-tailed
      preference distributions are associated with gradual polarization (see Panel~\ref{fig:bifurc-gg}).

    \item \textit{Cauchy kernel, Gaussian or Cauchy density.} When the
      satisficing kernel is Cauchy---that is, $f(x)=(1+x^{2})^{-1}$---and the voter
      density $\rho(x)$ is either Gaussian or itself Cauchy, the reduced
      equation exhibits pitchfork bifurcations at explicit critical values
      $q_{c}\approx 0.9753$ (Gaussian density) and $q_{c}\approx 0.9068$ (Cauchy
      density); see Fig.~\ref{fig:bifurc-cg}.

    \item \textit{Asymmetric two--group Gaussian electorate.} If the symmetry
      assumption on $\rho$ is dropped and the electorate is represented as an
      equally weighted mixture of two Gaussian components with means
      $(d_{1},d_{2})$ and standard deviations $(\sigma_{1},\sigma_{2})$, the symmetric
      pitchfork is unfolded. The centrist branch shifts away from the origin, a polarized
      branch appears, and the equilibrium diagram exhibits a hysteresis loop
      with lock--in of polarization as $q$ varies (see Section~\ref{sec:asymexamples}
      and Fig.~\ref{fig:gaussian-hysteresis}).
  \end{itemize}

  \section{Proof of the General Case}
  \label{sec:general}

  \subsection{First-Order Conditions in center-distance Coordinates}

  A Nash equilibrium is a pair $(u,v)$ solving the coupled system
  \[
    \partial_{u}V_{1,q}(u,v)=0 \quad\text{and}\quad \partial_{v}V_{2,q}(v,u)=0.
  \]
  To decouple the system into components representing the political center and
  the degree of polarization, we introduce the change of variables
  \[
    m \;=\; \frac{u+v}{2}, \qquad d \;=\; \frac{u-v}{2},
  \]
  where $m$ is the midpoint of the party positions and $d$ is half of the ideological
  distance between them. In these coordinates, the first--order conditions for a
  Nash equilibrium correspond to finding the zeros of the \emph{FOC vector field}
  \[
    \widehat F_{q}(m,d) :=\bigl(F_{q}^{+}(m,d),\,F_{q}^{-}(m,d)\bigr),
  \]
  where the symmetric and antisymmetric components are given by
  \begin{align}
    \label{eq:Gpm}F_{q}^{+}(m,d) & :=\frac{1}{2}\Bigl[\partial_{u}V_{1,q}+ \partial_{v}V_{2,q}\Bigr]_{(u,v)=(m+d,m-d)}, \\
    F_{q}^{-}(m,d)               & :=\frac{1}{2}\Bigl[\partial_{u}V_{1,q}- \partial_{v}V_{2,q}\Bigr]_{(u,v)=(m+d,m-d)}.
  \end{align}
 Thus, a pair $(u,v)$ satisfies the first--order conditions if and
only if
\[
  \widehat F_{q}\Bigl(\tfrac{u+v}{2},\,\tfrac{u-v}{2}\Bigr)=(0,0).
\]

  \begin{lemma}[Symmetry and Jacobian Structure]
  \label{lem:parity} \label{lem:triangular}
  For every fixed $m$, the components satisfy the parity relations
  \[
    F_{q}^{+}(m,-d)=F_{q}^{+}(m,d) \quad\text{and}\quad F_{q}^{-}(m,-d)=-\,F_{q}^{-}(m,d).
  \]
  Consequently, the off-diagonal derivatives vanish at $d=0$, that is, $\partial_{d}F_{q}^{+}(m,0)=0$ and $\partial_{m}F_{q}^{-}(m,0)=0$.
  In particular, at a center $m_{0}(q)$ defined by $F_{q}^{+}(m_{0}(q),0)=0$, the Jacobian takes the diagonal form
  \[
    D\widehat F_{q}(m_{0}(q),0)=
    \begin{pmatrix}
      \alpha(q) & 0      \\[2pt]
      0         & \mu(q)
    \end{pmatrix},
  \]
  where $\alpha(q) :=\partial_{m}F_{q}^{+}(m,0)\big|_{m=m_0(q)}$ and $\mu(q) :=\partial_{d}F_{q}^{-}(m,0)\big|_{m=m_0(q)}$.
\end{lemma}


  \begin{lemma}[Center exists for almost every $q$]
    \label{lem:center-exists} Fix $q>0$ and define
    \[
      \mathcal{G}(q,m):=F_{q}^{+}(m,0).
    \]
    Then there exists $m^{\ast}\in\mathbb{R}$ such that $\mathcal{G}(q,m^{\ast})=
    0$. Moreover, for almost every $q>0$ (in the Lebesgue measure sense), there
    is a zero $m^{\ast}(q)$ with $\partial_{m}\mathcal{G}(q,m^{\ast}(q))\neq0$.
  \end{lemma}

  \begin{proof}
    Differentiating the vote share $V_{1,q}(u,v)$ with respect to $u$ and applying
    the chain rule, we obtain
    \[
      \partial_{u} V_{1,q}(u,v) = \int_{\mathbb{R}}\left[-\frac{1}{q}\, f'\!\left
      (\tfrac{|x-u|}{q}\right)\operatorname{sgn}(x-u)\right] \left[ 1 - \frac{1}{2}
      f\!\left(\tfrac{|x-v|}{q}\right) \right] \rho(x) \, dx.
    \]
    Evaluating at the symmetric configuration $u=v=m$, and using $-\operatorname{sgn}
    (x-m)=\operatorname{sgn}(m-x)$, gives
    \[
      \partial_{u} V_{1,q}(m,m) = \int_{\mathbb{R}}\underbrace{ \frac{1}{q}\, f'\!\left(\tfrac{|x-m|}{q}\right)
      \Bigl[1-\frac{1}{2}f\!\left(\tfrac{|x-m|}{q}\right)\Bigr]
      \operatorname{sgn}(m-x) }_{=\,W_q(m-x)}\rho(x) \, dx,
    \]
    which is precisely the convolution $(W_{q} * \rho)(m)$.

    For $d=0$ we therefore have the convolution representation
    \[
      \mathcal{G}(q,m) = F_{q}^{+}(m,0) = \int_{\mathbb{R}}W_{q}(m-x)\,\rho(x)\,d
      x = (W_{q} * \rho)(m),
    \]
    where
    \[
      W_{q}(z) := w_{q}(|z|)\,\operatorname{sgn}z, \quad w_{q}(r) := \frac{1}{q}
      \,f'\,\!\Big(\tfrac{r}{q}\Big) \Bigl(1-\tfrac{1}{2}f\,\!\Big(\tfrac{r}{q}\Big)\Bigr
      ) \ \le 0
    \]
    for $r\ge 0$.

    Since $0\le 1-f\le 1$ and $f'\in L^{1}(0,\infty)$, the change of variables
    $t=r/q$ gives
\begin{align*}
      \int_{0}^{\infty}w_{q}(r)\,dr& =\int_{0}^{\infty}\bigl(-f'(t)\bigr)\bigl(1-f
      (t)\bigr)\,dt \\
      &\le \int_{0}^{\infty}\bigl(-f'(t)\bigr)\,dt = f(0)-\lim_{t\to\infty}
      f(t)=1,
\end{align*}
    so $w_{q}\in L^{1}(\mathbb{R}_{+})$ and hence $W_{q}\in L^{1}(\mathbb{R})$; also
    $W_{q}$ is odd. With $\rho\in L^{1}\cap L^{\infty}$, the convolution $\mathcal{G}
    (\cdot)=W_{q}*\rho$ is continuous and integrable.

    Because $W_{q}$ is odd and integrable while $\int_{\mathbb{R}}\rho=1$,
    Fubini’s theorem yields
   \begin{align*}
      \int_{\mathbb{R}}\mathcal{G}(q,m)\,dm &=\int_{\mathbb{R}}\!\!\int_{\mathbb{R}}
      W_{q}(m-x)\,\rho(x)\,dx\,dm \\
      &=\left(\int_{\mathbb{R}}W_{q}(y)\,dy\right)\!\left
      (\int_{\mathbb{R}}\rho(x)\,dx\right) =0.
    \end{align*}
    If $\mathcal{G}(q,\cdot)\equiv 0$ then every $m$ is a zero. Otherwise, since
    $\mathcal{G}(q,\cdot)$ is continuous and has zero integral, it must take both
    positive and negative values; by the Intermediate Value Theorem there exists
    $m^{\ast}$ with $\mathcal{G}(q,m^{\ast})=0$.

    \smallskip
    Since $\mathcal{G}$ is $C^{2}$ in $(q,m)$, we may view $\{H_{q}\}_{q>0}$ as a
    $C^{2}$ one–parameter family of maps $H_{q}:\mathbb{R}\to\mathbb{R}$,
    $H_{q}(m)=\mathcal{G}(q,m)$. By the parametric transversality theorem (see, e.g.,
    \cite[Chapter~2]{GuilleminPollack}), the set
    \begin{align*}
      \mathcal{T}&:=\{\,q>0:\ 0 \ \text{is a regular value of }H_{q}\,\} \\
      &=\{\,q>0:
      \ \partial_{m}\mathcal{G}(q,m)\neq 0 \ \text{whenever }\mathcal{G}(q,m)=0\,
      \}
\end{align*}
    has full Lebesgue measure in $(0,\infty)$ (equivalently, its complement has measure
    zero). Hence, for almost every $q>0$, all zeros of $m\mapsto \mathcal{G}(q,m)$
    are nondegenerate. Combining this with the existence of at least one zero
    for each $q>0$ proved above, we obtain that for almost every $q>0$ there exists
    $m^{\ast}(q)$ such that $\mathcal{G}(q,m^{\ast}(q))=0$ and $\partial_{m}\mathcal{G}
    (q,m^{\ast}(q))\neq 0$.
  \end{proof}

  \begin{lemma}[Local existence/uniqueness and regularity of the center]
    \label{lem:local-m0} Let $\mathcal{G}(q,m):=F_{q}^{+}(m,0)$ and suppose
    \[
      \mathcal{G}(q_{c},m^{\ast})=0 \quad \text{and}\quad \partial_{m}\mathcal{G}
      (q_{c}, m^{\ast})\neq 0.
    \]
    Then there exist $\delta>0$ and a unique $C^{1}$ function
    \[
      m_{0}:(q_{c}-\delta,q_{c}+\delta)\to\mathbb{R}
    \]
    such that $\mathcal{G}\bigl(q,m_{0}(q)\bigr)=0$ for all $q$ in that interval.
    Moreover,
    \[
      m_{0}'(q_{c}) \;=\; -\,\frac{\partial_{q}\mathcal{G}(q_{c},m^{\ast})}{\partial_{m}\mathcal{G}(q_{c},m^{\ast})}
      .
    \]
    If $\mathcal{G}\in C^{r}$ for some $r\ge 1$, then $m_{0}\in C^{r}$ (possibly
    after shrinking $\delta$).
  \end{lemma}

  \begin{proof}

    By the Implicit Function Theorem there exist $\delta>0$ and a unique $C^1$ map $m_0(q)$ with $\mathcal{G}(q,m_0(q))=0$ for $|q-q_c|<\delta$, and differentiating the identity yields the stated formula for $m_0'(q_c)$.

  \end{proof}

  \begin{remark}
    In our specific context, the assumption $f \in C^4$ implies $V \in C^4$, which entails $\mathcal{G} \in C^3$. Thus, the center branch $m_0(q)$ is guaranteed to be $C^3$.
  \end{remark}

  \begin{definition}[Center and linear coefficients]
    \label{def:center} For each $q>0$, a \emph{center} is any $m_{0}(q)$ solving
    \[
      F_{q}^{+}(m_{0}(q),0)=0.
    \]
    When multiple solutions exist, we work \emph{locally} near a nondegenerate point
    $(q_{c},m^{\ast})$ with $\mathcal{G}(q_{c},m^{\ast})=0$ and $\partial_{m}
    \mathcal{G}(q_{c},m^{\ast})\neq0$: by Lemma~\ref{lem:local-m0} there are
    $\delta>0$ and a unique $C^{1}$ map
    $m_{0}:(q_{c}-\delta,q_{c}+\delta)\to\mathbb{R}$ such that
    $\mathcal{G}(q,m_{0}(q))=0$ for all $q$ in that interval, with
    \[
      m_{0}'(q_{c}) =-\frac{\partial_{q}\mathcal{G}(q_{c},m^{\ast})}{\partial_{m}\mathcal{G}(q_{c},m^{\ast})}
      .
    \]
  \end{definition}

  \noindent
  \textit{Convention.} On any interval where a center branch $q\mapsto m_{0}(q)$
  is defined, we set
  \[
    \mu(q):=\partial_{d}F_{q}^{-}(m,0)\big|_{m=m_0(q)},\qquad \alpha(q):=\partial
    _{m}F_{q}^{+}(m,0)\big|_{m=m_0(q)}.
  \]

  \begin{lemma}[Continuity, representation, and sign change of $\mu$]
    \label{lem:mu-sign-change} Fix an interval $I\subset(0,\infty)$ and a continuous
    center branch $q\mapsto m_{0}(q)$ on $I$ as in Definition~\ref{def:center}.
    Then:
    \begin{enumerate}
      \item[(a)] For every $q\in I$,
        \[
          \mu(q) = \partial_{d}F_{q}^{-}(m,0)\Big|_{m=m_0(q)}= \frac{2}{q}\int_{0}
          ^{\infty}\psi(t)\,\tilde \rho_{q,e}(qt)\,dt,
        \]
        where
        \[
          \psi(t) := f''(t)\Bigl(1-\tfrac12 f(t)\Bigr) + \tfrac12\bigl(f'(t)\bigr
          )^{2}, \quad \tilde \rho_{q}(x):=\rho\bigl(m_{0}(q)+x\bigr),\]\[ \tilde
          \rho_{q,e}(x):=\tfrac12\bigl(\tilde \rho_{q}(x)+\tilde \rho_{q}(-x)\bigr
          ).
        \]

      \item[(b)] The map $q\mapsto\mu(q)$ is continuous on $I$.

      \item[(c)] There exist $0<q^{-}<q^{+}<\infty$ (depending only on $f$, $\rho$,
        and on the chosen branch $m_{0}$) such that, possibly after shrinking
        $I$,
        \[
          0<q<q^{-}\;\Longrightarrow\;\mu(q)>0, \qquad q>q^{+}\;\Longrightarrow\;
          \mu(q)<0.
        \]
        In particular, along this branch there exists $q_{c}\in(0,\infty)$
        with $\mu(q_{c})=0$.
    \end{enumerate}
  \end{lemma}

  \begin{proof}[Proof of Lemma~\ref{lem:mu-sign-change}]
By definition,
\[
  \mu(q)=\partial_{d}F_{q}^{-}(m,0)\Big|_{m=m_0(q)}.
\]
Differentiating the $d$–odd field $F_q^-(m,d)$ at $d=0$, using the scaled
kernel $f_q(r):=f(r/q)$ and the change of variables $x=qt$, we obtain
\[
  \mu(q) = \frac{2}{q}\int_{0}^{\infty}\psi(t)\,\tilde \rho_{q,e}(qt)\,dt,
\]
where
\[
  \psi(t) := f''(t)\Bigl(1-\tfrac12 f(t)\Bigr) + \tfrac12\bigl(f'(t)\bigr)^{2},
  \quad
  \tilde \rho_{q}(x):=\rho\bigl(m_{0}(q)+x\bigr),
\]
\[
  \tilde \rho_{q,e}(x):=\tfrac12\bigl(\tilde \rho_{q}(x)+\tilde \rho_{q}(-x)\bigr).
\]
Differentiation under the integral sign is justified by
Assumptions~\ref{assup1}–\ref{assup2}, since $f',f''\in L^{1}(0,\infty)$
and $\rho\in L^{1}\cap L^{\infty}$.

From $f(0)=1$, $f'(0)=0$, $f''(0)<0$ and $0\le f\le1$, a standard
integration–by–parts argument (using $f',f''\in L^{1}$ and
$f'\in L^{2}$) yields
\[
  \psi\in L^{1}(0,\infty),\quad
  \int_{0}^{\infty}\psi(t)\,dt=\int_{0}^{\infty}\bigl(f'(t)\bigr)^{2}\,dt>0,
  \]\[
  \psi(0)=\frac{f''(0)}{2}<0,
\]
and in particular $\psi$ is bounded on $[0,\infty)$.

The continuity of $\mu$ on $I$ follows directly from this representation:
if $q_n\to q_0\in I$, then
\[
  \tilde \rho_{q_n,e}(q_n t)
  = \tfrac12\bigl[\rho(m_0(q_n)+q_n t)+\rho(m_0(q_n)-q_n t)\bigr]
  \longrightarrow \tilde \rho_{q_0,e}(q_0 t)
\]
for each $t\ge0$, by continuity of $\rho$ and $m_0$, and
$|\tilde\rho_{q,e}(qt)|\le\|\rho\|_{L^\infty}$. Dominated convergence in
the integral for $\mu(q)$ then gives $\mu(q_n)\to\mu(q_0)$.

For small $q$, fix a compact subinterval $J_{-}\subset I$ accumulating at $0$,
and assume that
\[
  K_{-}:=\{m_{0}(q):q\in J_{-}\}
\]
lies in a compact set where $\rho$ is continuous and strictly positive.
Then $\rho(x)\ge c_{0}>0$ on $K_{-}$ for some $c_0$. By uniform continuity
of $\rho$ on $K_{-}$ and the fact that $qt\to0$ when $q\to0$ with $t$
bounded, we have
\[
  \tilde\rho_{q,e}(qt) = \rho(m_0(q)) + o(1)
\]
uniformly for $t$ in compact sets as $q\to0$. On the other hand, since
$\psi\in L^{1}$, we can truncate the integral at a large $T$ and make the
tail $\int_T^\infty|\psi(t)|dt$ arbitrarily small. Combining these two
facts in the representation of $q\,\mu(q)$ gives
\[
  q\,\mu(q)
  = 2\int_{0}^{\infty}\psi(t)\,\tilde\rho_{q,e}(qt)\,dt
  = 2\rho(m_0(q))\int_{0}^{\infty}\psi(t)\,dt + o(1)
\]
as $q\to0$. Since $\rho(m_0(q))\ge c_0$ and
\[
  \int_{0}^{\infty}\psi(t)\,dt
  = \int_{0}^{\infty}\bigl(f'(t)\bigr)^{2}\,dt>0,
\]
it follows that $q\,\mu(q)>0$ for $q>0$ small enough, hence
$\mu(q)>0$ for $0<q<q^{-}$ for some $q^{-}>0$.

For large $q$, take a ray $J_{+}\subset I$ with $\inf J_{+}>0$ and suppose
\[
  K_{+}:=\{m_0(q):q\in J_{+}\}\subset[-R_{+},R_{+}]
\]
for some $R_+<\infty$ (after possibly shrinking $I$). Fix $\varepsilon>0$
and choose $M>0$ such that
\[
  \int_{|x|>M}\rho(x)\,dx<\varepsilon,
\]
and set $T_q:=(M+R_+)/q$. Splitting the integral,
\begin{align*}
  \mu(q)
  &= \frac{2}{q}\int_{0}^{T_q}\psi(t)\,\tilde \rho_{q,e}(qt)\,dt
    + \frac{2}{q}\int_{T_q}^{\infty}\psi(t)\,\tilde \rho_{q,e}(qt)\,dt \\
  &=: J_1(q)+J_2(q),
\end{align*}
we first bound the tail. With $y=qt$,
\[
  \int_{T_q}^{\infty}\tilde \rho_{q,e}(qt)\,dt
  = \frac{1}{q}\int_{M+R_+}^{\infty}\tilde\rho_{q,e}(y)\,dy\le \int_{|z|>M}\rho(z)\,dz < \varepsilon,
\]
uniformly in $q\in J_+$. Using $|\psi|\le\|\psi\|_\infty$ gives
$|J_2(q)|\le C\,\varepsilon/q^{2}$.

On $[0,T_q]$ we have $qt\le M+R_+$ and hence $m_0(q)\pm qt$ remain in a
compact set where $\rho$ is bounded. Moreover, $T_q\to0$ as $q\to\infty$,
so by continuity of $\psi$ at $0$ we have $\psi(t)=\psi(0)+o(1)$ uniformly
for $0\le t\le T_q$. Writing
\[
  J_1(q)
  = \frac{2\psi(0)}{q}\int_{0}^{T_q}\tilde\rho_{q,e}(qt)\,dt
    + \frac{2}{q}\int_{0}^{T_q}(\psi(t)-\psi(0))\,\tilde\rho_{q,e}(qt)\,dt,
\]
and changing variables $y=qt$ in the first term, we obtain
\[
  \frac{2\psi(0)}{q}\int_{0}^{T_q}\tilde\rho_{q,e}(qt)\,dt
  = \frac{2\psi(0)}{q^{2}}\int_{0}^{M+R_+}\tilde\rho_{q,e}(y)\,dy.
\]
A simple change–of–variables argument shows that
\[
  \int_{0}^{M+R_+}\tilde\rho_{q,e}(y)\,dy
  \ge \frac{1}{2}\int_{-M}^{M}\rho(z)\,dz =: \mu_0>0
\]
uniformly in $q\in J_+$. The second term in $J_1(q)$ is of order
$O(\varepsilon/q^{2})$ by the smallness of $T_q$ and the boundedness of
$\tilde\rho_{q,e}$. Thus
\[
  \mu(q)
  \le \frac{2\psi(0)\mu_0}{q^{2}} + \frac{C\varepsilon}{q^{2}}
\]
for all $q\in J_+$, with $\psi(0)=f''(0)/2<0$. Choosing $\varepsilon$ small
and then $q$ large we ensure that the right–hand side is negative, so there
exists $q^{+}>0$ such that $\mu(q)<0$ for all $q>q^{+}$ in $I$.

We have shown that $\mu(q)>0$ for $0<q<q^{-}$, $\mu(q)<0$ for $q>q^{+}$,
and $\mu$ is continuous on $I$. By the Intermediate Value Theorem there
exists $q_{c}\in(0,\infty)$ such that $\mu(q_{c})=0$.
\end{proof}

  \begin{corollary}[Existence of a crossing]
    \label{cor:mu-crossing} Under Lemma~\ref{lem:mu-sign-change}, there exists
    $q_{c}\in(0,\infty)$ such that $\mu(q_{c})=0$.
  \end{corollary}

  \subsection{Local Pitchfork Bifurcation}

  \begin{definition}[Effective Cubic Coefficient]
    \label{def:beta-eff} For a $C^{3}$ center branch $m_{0}(q)$, we define the
    \emph{effective cubic coefficient} $\beta_{\mathrm{eff}}(q)$ as the coefficient
    that incorporates the pure cubic term, the branch curvature, and the mixed-derivative
    coupling:
    \[
      \beta_{\mathrm{eff}}(q):=\beta(q)+\tfrac12\,C_{md}(q)\,h_{dd}(q,0),
    \]
    where the components are defined along the branch $m_{0}(q)$ as:
    \begin{itemize}
      \item $\beta(q):=\tfrac16\,\partial_{ddd}F_{q}^{-}(m_{0}(q),0)$ (the pure
        cubic term);

      \item $C_{md}(q):=\partial_{md}F_{q}^{-}(m_{0}(q),0)$ (the coupling term);

      \item $h_{dd}(q,0)$ (branch curvature), derived from the IFT map
        $m=m_{0}(q)+h(q,d)$ which solves $F_{q}^{+}(m_{0}(q)+h(q,d),d)=0$.
    \end{itemize}
  \end{definition}

  With this convention, $\beta_{\mathrm{eff}}
  (q_{c})<0$ yields a \emph{supercritical} pitchfork (stable side branches), whereas
  $\beta_{\mathrm{eff}}(q_{c})>0$ yields a \emph{subcritical} one.

  \begin{theorem}[Local pitchfork bifurcation]
    \label{thm:local-pitchfork-LS} Fix $q_{c}>0$ and assume
    \[
      \alpha(q_{c})\neq 0,\qquad \mu(q_{c})=0,\ \mu'(q_{c})\neq0,\qquad
      \beta_{\mathrm{eff}}(q_{c})<0.
    \]
    Then $(m,d)=(m_{0}(q),0)$ undergoes a \emph{supercritical pitchfork} at
    $q=q_{c}$ (\emph{i.e.}, side branches appear for $\mu(q)>0$). More precisely,
    there exist neighborhoods $I\ni q_{c}$ and
    $\mathcal{U}\ni(m_{0}(q_{c}),0)$ such that, for $q\in I$,
    \[
      \begin{cases}
        \mu(q)<0: & \text{the only equilibrium in }\mathcal{U}\text{ is }(m,d)=(m_{0}(q),0),                                               \\[3pt]
        \mu(q)>0: & \parbox{5cm}{\text{there are exactly three equilibria in }$\mathcal{U}:$\\ $(m_{0}(q),0)$ and two nontrivial branches,}
      \end{cases}
    \]
    with
    \[
      d_{\pm}(q)=\pm\sqrt{-\,\mu(q)/\beta_{\mathrm{eff}}(q)}+o\,\!\big(\sqrt{|\mu(q)|}
      \big),\]
      \[ m_{\pm}(q)=m_{0}(q)+h\bigl(q,d_{\pm}(q)\bigr)=m_{0}(q)+\mathcal{O}
      (\mu(q)).
    \]
  \end{theorem}

  \begin{proof}[Proof (Normal Form)]
    By Lemma~\ref{lem:local-m0} and Assumptions~\ref{assup1} and \ref{assup2},
    there exists $\delta>0$ and a unique center branch
    $m_{0}\in C^{3}((q_{c}-\delta,q_{c}+\delta))$ with
    $F_{q}^{+}(m_{0}(q),0)=0$. Set
    \[
      \alpha(q):=\partial_{m}F_{q}^{+}(m,0)\big|_{m=m_0(q)}.
    \]

    Since $F_{q}^{+}$ is even in $d$ (Lemma~\ref{lem:parity}), $\partial_{d}F_{q}
    ^{+}(m,0)=0$. Because $\alpha(q_{c})\neq0$, shrinking $I$ if needed there
    is $a_{0}>0$ with $|\alpha(q)|\ge a_{0}$ for $q\in I$. By the IFT, there is a
    unique $C^{3}$ map
    \[
      m=m_{0}(q)+h(q,d),\qquad F_{q}^{+}\bigl(m_{0}(q)+h(q,d),\,d\bigr)=0,
    \]
    with $h(q,0)=0$, $\partial_{d}h(q,0)=0$, and $h(q,d)=\mathcal{O}(d^{2})$ uniformly
    in $q\in I$. Differentiating twice in $d$ yields
    \[
      h_{dd}(q,0)=-\,\frac{F_{dd}^{+}(m_{0}(q),0)}{\alpha(q)}.
    \]

    \smallskip

    Define $\phi(q,d) :=F_{q}^{-}\!\bigl(m_{0}(q)+h(q,d),\,d\bigr)$. Using parity
    of $F_{q}^{-}$ and the $C^{3}$ regularity,
    \[
      \phi(q,d)=\mu(q)\,d + C_{md}(q)\,h(q,d)\,d + \beta(q)\,d^{3}+ r(q,d),
    \]
    with $r(q,d)=o(d^{3})$ uniformly in $q$. Since $h(q,d)=\tfrac12 h_{dd}(q,0)d^{2}
    +o(d^{2})$,
    \begin{align*}
      \phi(q,d)&=\mu(q)\,d+\Bigl[\beta(q)+\tfrac12\,C_{md}(q)\,h_{dd}(q,0)\Bigr]d^{3}
      +r(q,d)\\
      &=\mu(q)\,d+\beta_{\mathrm{eff}}(q)\,d^{3}+r(q,d),
\end{align*}
    and, by the formula for $h_{dd}$,
    \[
      \beta_{\mathrm{eff}}(q) =\beta(q)-\tfrac12\,C_{md}(q)\,\frac{F_{dd}^{+}(m_{0}(q),0)}{\alpha(q)}
      .
    \]
    By continuity, $\beta_{\mathrm{eff}}(q_{c})<0$ implies (after shrinking
    $I$) that $\beta_{\mathrm{eff}}(q)\le -\beta_{\ast}<0$ for $q\in I$.

    \smallskip
    Thus
    \[
      \phi(q,d)=d\bigl(\mu(q)+\beta_{\mathrm{eff}}(q)\,d^{2}+o(d^{2})\bigr).
    \]
    If $\mu(q)<0$, the bracket is negative for small $|d|$, giving only $d=0$.
    If $\mu(q)>0$, there are exactly two nonzero small roots
    \[
      d_{\pm}(q)=\pm\sqrt{-\,\mu(q)/\beta_{\mathrm{eff}}(q)}+o\,\!\big(\sqrt{|\mu(q)|}
      \big),
    \]
    and $m=m_{0}(q)+h(q,d_{\pm}(q))=m_{0}(q)+\mathcal{O}(\mu(q))$. By the graph
    property, every solution of $F_{q}^{+}=0$ in a sufficiently small neighborhood
    satisfies $m = m_{0}(q) + h(q,d)$, so the equilibria of the full system in
    $\mathcal{U}$ are precisely the zeros of $\phi(q,\cdot)$ listed above.
  \end{proof}

  \begin{remark}[Hysteresis]
    In the symmetric benchmark, a supercritical pitchfork yields a smooth and
    reversible transition from convergence to polarization, with a unique locally
    stable equilibrium for each value of $q$, so hysteresis cannot occur. Once asymmetry
    is introduced, the pitchfork is imperfectly unfolded and the equilibrium
    curve in $(q,d)$ can become S-shaped, leading to hysteresis.
  \end{remark}

  \section{Proof of the high-tolerance limit: convergence to the mean}
  \label{sec:highq-proof}
  The next lemma shows that, for all sufficiently large $q$,
  each player’s objective is uniformly strictly concave on a common interval $K$
  and has a unique global maximizer in $K$, independently of the opponent’s position.

  \begin{lemma}[Uniqueness of the maximizer for large tolerance]
\label{lem:unique-max-large-q}
Suppose Assumptions~\ref{assup1}--\ref{assup3} hold, and assume in addition that
\[
  f'' \in L^{1}(0,\infty)\cap L^{\infty}(0,\infty)
  \quad\text{and}\quad
  \int_{\mathbb{R}} |x|\,\rho(x)\,dx < \infty .
\]
Then there exist a constant \(Q>0\) and a compact interval \(K\subset\mathbb{R}\),
depending only on \((f,\rho)\), such that the following holds.
\begin{enumerate}
  \item[(a)] For every \(q\ge Q\), every player \(i\in\{1,2\}\), and every fixed
  opponent position \(u_j\in\mathbb{R}\), the objective
  \[
    u \longmapsto V_{i,q}(u,u_j)
  \]
  is strictly concave on \(K\).

  \item[(b)] For every \(q\ge Q\), every global maximizer  of
  \(V_{i,q}(\,\cdot\,,u_j)\) belongs to \(K\), and this global maximizer is unique.
\end{enumerate}
\end{lemma}

\begin{proof}

  Fix $i\in\{1,2\}$ and $u_{j}\in\mathbb{R}$, and write
  \[
    w_{q,u_j}(x) := \Bigl[1 - \tfrac12 f_{q}(|x-u_{j}|)\Bigr]\rho(x),
  \]
  so that
  \begin{equation}
    \label{eq:w-bounds}
    \tfrac12 \rho(x) \;\le\; w_{q,u_j}(x) \;\le\; \rho(x) \qquad
    \text{for all }x\in\mathbb{R},\ q>0.
  \end{equation}
  Then
  \[
    V_{i,q}(u,u_{j}) = \int_{\mathbb{R}}f\left(\frac{|x-u|}{q}\right)\,w_{q,u_j}(x)\,dx.
  \]

  For each fixed $x$ and $q$, define
  \[
    g_{x}(u) := f\left(\frac{|x-u|}{q}\right).
  \]
  A direct computation gives, for $u\neq x$,
  \[
    \partial_{u}^{2} g_{x}(u) = \frac{1}{q^{2}}\,f''\left(\frac{|x-u|}{q}\right),
  \]
  and, by dominated convergence (using $f''\in L^{\infty}$ and
  $w_{q,u_j}\in L^{1}$),
  \begin{equation}
    \label{eq:Vpp-short}
    \partial_{u}^{2} V_{i,q}(u,u_{j}) = \frac{1}{q^{2}}\int_{\mathbb{R}}
    f''\left(\frac{|x-u|}{q}\right)\,w_{q,u_j}(x)\,dx.
  \end{equation}

  \medskip
  \noindent
  \emph{Uniform strict concavity on a fixed interval.}
  Since $f''(0)<0$ and $f''$ is continuous at $0$, there exist $\delta>0$
  and $\gamma>0$ such that
  \[
    f''(t) \le -\gamma \quad\text{for all }t\in[0,\delta].
  \]
  Because $\rho$ is integrable, we can choose $R_{0}>0$ so that
  \[
    \int_{|x|>R_0}\rho(x)\,dx < \varepsilon
  \]
  for some small $\varepsilon>0$ to be fixed. Using \eqref{eq:w-bounds}, this
  implies that a positive fraction of the mass of $w_{q,u_j}$ lies in
  $[-R_{0},R_{0}]$, uniformly in $q$ and $u_{j}$, while the total mass of
  $w_{q,u_j}$ outside $[-R_{0},R_{0}]$ is at most $\varepsilon$.

  Choose $Q_{1}>0$ such that $\delta Q_{1}>2R_{0}$. Then, for every
  $q\ge Q_{1}$, every $u\in[-R_{0},R_{0}]$, and every $x\in[-R_{0},R_{0}]$
  we have $|x-u|/q\le\delta$, hence $f''(|x-u|/q)\le -\gamma$ on
  $[-R_{0},R_{0}]$. Splitting the integral in~\eqref{eq:Vpp-short} into
  $|x|\le R_{0}$ and $|x|>R_{0}$, we obtain
  \[
    q^{2}\,\partial_{u}^{2} V_{i,q}(u,u_{j}) \le -\gamma\,c_{0} + M\varepsilon,
  \]
  where $c_{0}>0$ depends only on $\rho$ (via the mass of $w_{q,u_j}$ in
  $[-R_{0},R_{0}]$) and $M:=\sup_{t\ge0}|f''(t)|<\infty$. Choosing
  $\varepsilon>0$ small enough so that $M\varepsilon<\gamma c_{0}/2$, we find
  a constant $\kappa>0$ such that
  \[
    \partial_{u}^{2} V_{i,q}(u,u_{j}) \le -\frac{\kappa}{q^{2}}< 0 \quad
    \text{for all }u\in[-R_{0},R_{0}],\ q\ge Q_{1},\ u_{j}\in\mathbb{R}.
  \]
  Thus $V_{i,q}(\cdot,u_{j})$ is strictly concave on
  $K_{0}:=[-R_{0},R_{0}]$ for all $q\ge Q_{1}$.

  \medskip
  \noindent
  \emph{Uniform boundedness and uniqueness of maximizers.}
  For each $q>0$ and $u_{j}$, $V_{i,q}(\cdot,u_{j})$ attains at least one
  global maximum on $\mathbb{R}$.

  Next, using $f(0)=1$, $f'(0)=0$, $f''(0)<0$ and the integrability/boundedness
  assumptions on $f''$, a standard Taylor expansion under the integral sign yields
  constants $Q_{2}>0$ and $C'>0$ such that, for all $q\ge Q_{2}$ and all
  $u,u_{j}\in\mathbb{R}$,
  \begin{equation}
    \label{eq:V-expansion-short}
    V_{i,q}(u,u_{j}) = A(q,u_{j}) + \frac{f''(0)}{2q^{2}}\,J_{w}(u) +
    R_{q}(u,u_{j}),
  \end{equation}
  where $A(q,u_{j})$ is independent of $u$,
  \[
    J_{w}(u) := \int_{\mathbb{R}}(x-u)^{2}\,w_{q,u_j}(x)\,dx
  \]
  is strictly convex with a unique minimizer, which we denote by $u_{\mathrm{b}}$,
  and the remainder satisfies $|R_{q}(u,u_{j})|\le C'/q^{3}$ uniformly in
  $u,u_{j}$.

  The minimizer $u_{\mathrm{b}}$ is the barycenter of $w_{q,u_j}$:
  \[
    u_{\mathrm{b}} =
    \frac{\int_{\mathbb{R}}x\,w_{q,u_j}(x)\,dx}
         {\int_{\mathbb{R}}w_{q,u_j}(x)\,dx}.
  \]
  Using \eqref{eq:w-bounds} and the finite first moment of $\rho$, we obtain a
  uniform bound $|u_{\mathrm{b}}|\le C_{1}$ for some $C_{1}>0$ depending only
  on $\rho$. Moreover $J_{w}(u)-J_{w}(u_{\mathrm{b}})\ge c_{1}>0$ whenever
  $|u|\ge R_{1}$ for a suitably large $R_{1}>0$ (e.g.\ $R_{1}>2C_{1}$), by
  strict convexity and \eqref{eq:w-bounds}. Plugging this into
  \eqref{eq:V-expansion-short}, we find $Q_{3}\ge Q_{2}$ such that, for all
  $q\ge Q_{3}$,
  \[
    V_{i,q}(u,u_{j}) < V_{i,q}(u_{\mathrm{b}},u_{j}) \qquad
    \text{whenever }|u|\ge R_{1}.
  \]
  Hence every global maximizer lies in
  \[
    K' := [-R_{1},R_{1}],
  \]
  which depends only on $(f,\rho)$ and is independent of $q$ and $u_{j}$.

  Finally, let
  \[
    K := [-R,R], \qquad R := \max\{R_{0},R_{1}\}, \qquad
    Q := \max\{Q_{1},Q_{3}\}.
  \]
  For every $q\ge Q$ and every $u_{j}$, any global maximizer of
  $V_{i,q}(\cdot,u_{j})$ belongs to $K'\subset K$, and $V_{i,q}(\cdot,u_{j})$
  is strictly concave on the convex set $K$. A strictly concave function on a
  convex interval has at most one global maximizer in that interval, so the
  global maximizer is unique and lies in $K$, as claimed.

\end{proof}

  Under the nondegeneracy assumptions of Theorem~\ref{thm:local-pitchfork-LS},
  the loss of stability of the symmetric equilibrium along this branch occurs
  through a $\mathbb{Z}_{2}$–equivariant pitchfork bifurcation in the polarization
  coordinate $d$. In particular, for $q>q_{c}$ there is a unique locally stable
  symmetric equilibrium, while for $q<q_{c}$ sufficiently close to $q_{c}$
  this equilibrium becomes unstable and two locally stable, oppositely polarized
  equilibria emerge. In particular, generically the bifurcation at $q=q_{c}$ is
  a supercritical pitchfork in $d$.

  While Lemma~\ref{lem:unique-max-large-q} guarantees that best replies are well
  behaved and unique for $q$ large, the next lemma shows that these maximizers
  converge to the mean $\bar x$ of the voter distribution as $q\to\infty$, independently
  of the opponent’s position.

  \begin{lemma}[Large--$q$ limit: convergence to the mean ]
    \label{lem:large-q-limit} Suppose Assumptions~\ref{assup1}--\ref{assup3}
    hold and, in addition,
    \[
      \int_{\mathbb{R}}|x|^{2}\,\rho(x)\,dx < \infty.
    \]
    For each $i\in\{1,2\}$ and $u_{j}\in\mathbb{R}$, Lemma~\ref{lem:unique-max-large-q}
    provides a constant $Q_{0}>0$ such that, for all $q\ge Q_{0}$, the function $u
    \mapsto V_{i,q}(u,u_{j})$ has a unique global maximizer.
    Define the \textit{mean} of the voter density by
    \[
      \bar x := \int_{\mathbb{R}}x\,\rho(x)\,dx.
    \]
    Then, for every fixed $u_{j}\in\mathbb{R}$ and $i\in\{1,2\}$,
    \[
      u_{i}(q,u_{j}) \longrightarrow \bar x \qquad\text{as }q\to\infty.
    \]
    In particular, the limit is independent of $u_{j}$ and of the player $i$.
  \end{lemma}
\begin{proof}
Fix $i\in\{1,2\}$ and $u_{j}\in\mathbb{R}$. Recall that
\[
  V_{i,q}(u,u_{j})
  = \int_{\mathbb{R}}f\left(\frac{|x-u|}{q}\right)\,w_{q,u_j}(x)\,dx,
 \]\[
  w_{q,u_j}(x)
  := \Bigl[1 - \tfrac12 f_{q}(|x-u_{j}|)\Bigr]\rho(x),
\]
and that
\begin{equation}
  \tfrac12\rho(x) \le w_{q,u_j}(x) \le \rho(x)
  \qquad\text{for all }x\in\mathbb{R},\ q>0.
\end{equation}

Since $f_q(r)=f(r/q)$ and $f$ is continuous with $f(0)=1$, we have
$f_{q}(|x-u_j|)\to 1$ for every fixed $x$, hence
\[
  w_{q,u_j}(x) \to \tfrac12\rho(x)
  \qquad\text{for every }x\in\mathbb{R}.
\]
Moreover, $|w_{q,u_j}|\le\rho\in L^1$, so dominated convergence yields
\begin{equation}
  \label{eq:wq-conv}
  \int_{\mathbb{R}}\bigl|w_{q,u_j}(x) - \tfrac12\rho(x)\bigr|\,dx \;\longrightarrow\; 0
  \qquad\text{as }q\to\infty.
\end{equation}

For fixed $x$ and $q$, define $g_x(u):=f(|x-u|/q)$. A direct computation gives,
for $u\neq x$,
\[
  \partial_{u}^{2} g_{x}(u)
  = \frac{1}{q^{2}}\,f''\left(\frac{|x-u|}{q}\right),
\]
and, by dominated convergence (using $f''\in L^\infty$ and $w_{q,u_j}\in L^1$),
\begin{equation}
  \partial_{u}^{2} V_{i,q}(u,u_{j})
  = \frac{1}{q^{2}}\int_{\mathbb{R}}
      f''\left(\frac{|x-u|}{q}\right)\,w_{q,u_j}(x)\,dx.
\end{equation}

By Assumption~\ref{assup1}(c),
$
  f(0)=1,\,f'(0)=0,\,f''(0)<0,
$ 
and $f\in C^{3}(\mathbb{R}_{+})$. Thus, for each $t\ge 0$,
\[
  f(t) = 1 + \frac{1}{2} f''(0)t^{2} + r(t),
\]
where $r(t)=o(t^{2})$ as $t\to 0$. In particular, for every $\varepsilon>0$
there exists $\delta>0$ such that
\begin{equation}
  \label{eq:r-small}
  |r(t)| \le \varepsilon t^{2}
  \qquad\text{whenever }0\le t\le \delta.
\end{equation}
Since $f$ is bounded and $t\mapsto t^{2}$ has at most quadratic growth, there
exists $C_{\delta}>0$ such that
\begin{equation}
  \label{eq:r-large}
  |r(t)| \le C_{\delta} t^{2}
  \qquad\text{for all }t\ge \delta.
\end{equation}

Substituting $t = |x-u|/q$, we obtain
\begin{equation}
  \label{eq:fq-expansion}
  f_{q}(|x-u|)
  = f\left(\frac{|x-u|}{q}\right)
  = 1 + \frac{1}{2} f''(0)\,\frac{|x-u|^{2}}{q^{2}}+ r_{q}(x,u),
\end{equation}
with
\[
  r_{q}(x,u) := r\left(\frac{|x-u|}{q}\right),
\]
and, from \eqref{eq:r-small}–\eqref{eq:r-large},
\begin{equation}
  \label{eq:rq-bounds}
  |r_{q}(x,u)| \le
  \begin{cases}
    \varepsilon\,\dfrac{|x-u|^2}{q^2}, & \text{if }\dfrac{|x-u|}{q}\le\delta, \\[0.4em]
    C_{\delta}\,\dfrac{|x-u|^2}{q^2},  & \text{if }\dfrac{|x-u|}{q}\ge\delta.
  \end{cases}
\end{equation}

Fix a compact interval $K\subset\mathbb{R}$ containing all maximizers for
$q$ large, as given by Lemma~\ref{lem:unique-max-large-q}; enlarging $K$ if needed,
assume also $\bar x\in K$. For $u\in K$ and $q>0$, we write
\[
  V_{i,q}(u,u_{j})
  = A_{q}(u_{j})
    + \frac{1}{2} f''(0)\,\frac{1}{q^{2}}\int_{\mathbb{R}}|x-u|^{2}\,w_{q,u_j}(x)\,dx
    + R_{q}(u,u_{j}),
\]
where
\[
  A_{q}(u_{j}) := \int_{\mathbb{R}}w_{q,u_j}(x)\,dx
\]
does not depend on $u$, and
\[
  R_{q}(u,u_{j}) := \int_{\mathbb{R}}r_{q}(x,u)\,w_{q,u_j}(x)\,dx.
\]

Using \eqref{eq:rq-bounds}, the estimate $w_{q,u_j}\le\rho$ and the finite second moment of $\rho$, we split the integral defining $R_q$ into the regions $\{|x-u|\le\delta q\}$ and $\{|x-u|>\delta q\}$ and bound each part separately; this gives
\begin{equation}
  \label{eq:Rq-small-uniform}
  \sup_{u\in K}|q^{2} R_{q}(u,u_{j})| \;\longrightarrow\; 0
  \qquad\text{as }q\to\infty.
\end{equation}

We next show that
\begin{equation}
  \label{eq:second-moment-limit}
  \int_{\mathbb{R}}|x-u|^{2}\,w_{q,u_j}(x)\,dx
  \longrightarrow \frac{1}{2}\int_{\mathbb{R}}|x-u|^{2}\,\rho(x)\,dx
\end{equation}
as $q\to\infty$, uniformly in $u\in K$. Indeed, using
\eqref{eq:wq-conv} on a fixed ball $\{|x|\le R\}$ (where $|x-u|^{2}$ is
uniformly bounded for $u\in K$) and the finite second moment of $\rho$
to control the tail $\{|x|>R\}$, we obtain that the difference between
the two sides of \eqref{eq:second-moment-limit} converges to $0$
uniformly in $u\in K$.

Combining the decomposition of $V_{i,q}$ with \eqref{eq:Rq-small-uniform}
and \eqref{eq:second-moment-limit}, we can rewrite, for $u\in K$ and $q$ large,
\[
  V_{i,q}(u,u_{j})
  = A_{q}(u_{j})
    + \frac{f''(0)}{4q^{2}}\int_{\mathbb{R}}|x-u|^{2}\,\rho(x)\,dx
    + \widetilde R_{q}(u,u_{j}),
\]
where
\[
  \sup_{u\in K}|q^{2}\widetilde R_{q}(u,u_{j})| \;\longrightarrow\; 0
  \qquad\text{as }q\to\infty.
\]

Define
\[
  J(u) := \int_{\mathbb{R}}|x-u|^{2}\,\rho(x)\,dx.
\]
A standard computation (differentiating under the integral sign and using the
finite second moment of $\rho$) shows that $J$ is $C^2$, strictly convex, and
has a unique minimizer
\[
  \bar x = \int_{\mathbb{R}}x\,\rho(x)\,dx.
\]
Since $\bar x\in K$, the minimizer of $J$ over $K$ is also $\bar x$.

By Lemma~\ref{lem:unique-max-large-q}, for $q$ large the maximizer
$u_i^*(q,u_j)$ is unique and belongs to $K$. The uniform expansion of $V_{i,q}$ on $K$ and the strict convexity of $J$ then imply, by continuity of maximizers
under uniform perturbations of the objective, that
\[
  u_{i}(q,u_{j}) \longrightarrow \bar x \qquad\text{as }q\to\infty.
\]

\end{proof}

  In other words, when tolerance is very high the satisficing spatial competition
  effectively collapses to a quadratic loss minimization problem, and both
  parties optimally choose the mean of the voter density.

  \begin{corollary}[Symmetric benchmark: convergence to the median]
    \label{cor:median-limit} In addition to the hypotheses of Lemma~\ref{lem:large-q-limit},
    assume that $\rho$ is symmetric with respect to some $m\in\mathbb{R}$ and
    that $\rho$ has a unique median at $m$. Then the mean satisfies $\bar x = m$,
    and hence, for every fixed $u_{j}$,
    \[
      u_{i}(q,u_{j}) \longrightarrow m \qquad\text{as }q\to\infty.
    \]
    In particular, in the symmetric benchmark the unique equilibrium branch on the diagonal converges to the median voter as tolerance $q\to\infty$.
  \end{corollary}

  \section{Asymmetric Electorate Example}
  \label{sec:asymexamples}

  We now illustrate the previous analysis in a case with an asymmetric two--group
  electorate. We assume a Gaussian satisficing kernel:
  \[
    f(x) \;=\; e^{-x^{2}/2}.
  \]
  The voter density $\rho(x)$ is a mixture of two Gaussian groups:
  \[
    \rho(x) \;=\; \frac{1}{\sqrt{2\pi}\,(\sigma_{1}+\sigma_{2})}\left[e^{-(x+d_{1})^{2}/(2\sigma_{1}^{2})}
    \;+\; e^{-(x+d_{2})^{2}/(2\sigma_{2}^{2})}\right],
  \]
  with possibly different locations $d_{1},d_{2}$ and dispersions
  $\sigma_{1},\sigma_{2}$. This specification captures an electorate with two ideological
  blocs of unequal size and spread, shifting the center of gravity away from the
  origin and breaking the symmetry of the benchmark case.

  For this class of densities, the first--order conditions for party payoffs
  admit closed expressions after a suitable change of variables. Writing
  \[
    x \;=\; v-u, \qquad y \;=\; v+u,
  \]
  and introducing the shorthand
  \[
    A_{i}\;=\; q^{2}+\sigma_{i}^{2}, \qquad B_{i}\;=\; q^{2}+2\sigma_{i}^{2}, \qquad
    i\in\{1,2\},
  \]
  the partial derivatives of $V_{1,q}$ and $V_{2,q}$ can be written in the compact
  form
  \begin{align*}
    (\sigma_{1} +\sigma_{2})q^{2}\frac{\partial V_{1,q}}{\partial u}(u,v)  = -\sum_{i=1}^{2}\frac{q^{3}\sigma_{i}}{A_{i}^{3/2}}\,(u+d_{i}) \exp\!\left[-\frac{(u+d_{i})^{2}}{2A_{i}}\right]                                                                                \\
                                                                           \quad +\sum_{i=1}^{2}\left[ \frac{q^{3}\sigma_{i}}{4B_{i}^{3/2}}(y+2d_{i}) - \frac{q\sigma_{i}}{4\sqrt{B_{i}}}x \right] \exp\!\left[-\frac{(y+2d_{i})^{2}}{4B_{i}}-\frac{x^{2}}{4q^{2}}\right],
  \end{align*}
  \begin{align*}
    (\sigma_{1} +\sigma_{2})q^{2}\frac{\partial V_{2,q}}{\partial v}(v,u)  = -\sum_{i=1}^{2}\frac{q^{3}\sigma_{i}}{A_{i}^{3/2}}\,(v+d_{i}) \exp\!\left[-\frac{(v+d_{i})^{2}}{2A_{i}}\right]                                                                                \\
                                                                           \quad +\sum_{i=1}^{2}\left[ \frac{q^{3}\sigma_{i}}{4B_{i}^{3/2}}(y+2d_{i}) + \frac{q\sigma_{i}}{4\sqrt{B_{i}}}x \right] \exp\!\left[-\frac{(y+2d_{i})^{2}}{4B_{i}}-\frac{x^{2}}{4q^{2}}\right].
  \end{align*}
  Equilibrium party positions are therefore obtained by solving the system $\partial
  _{u} V_{1,q}(u,v)=0$ and $\partial_{v} V_{2,q}(v,u)=0$ as a function of the tolerance
  parameter $q$.

\begin{figure}[!h]
    \centering

    \begin{subfigure}
      {0.48\textwidth}
      \centering
      \includegraphics[width=\linewidth]{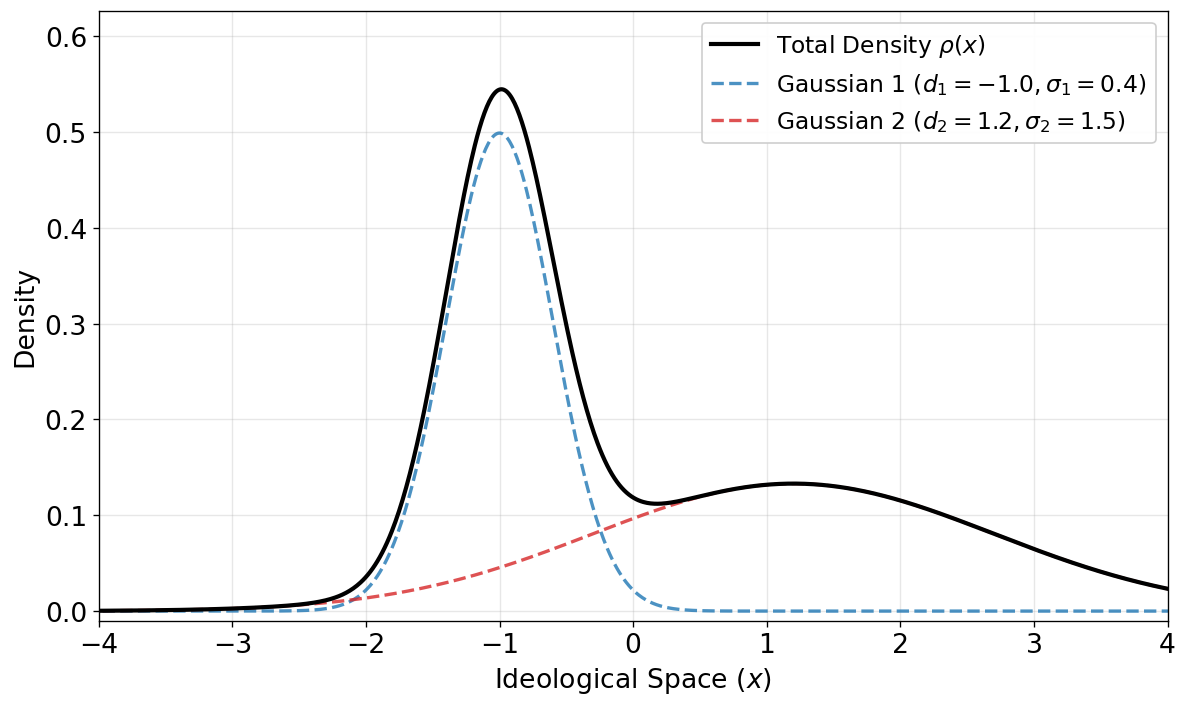}
      \caption{Voter distribution $\rho(x)$ (Gaussian mixture).}
      \label{fig:voter-gaussian-mixture}
    \end{subfigure}
    \hfill
    \begin{subfigure}
      {0.48\textwidth}
      \centering
      \includegraphics[width=\linewidth]{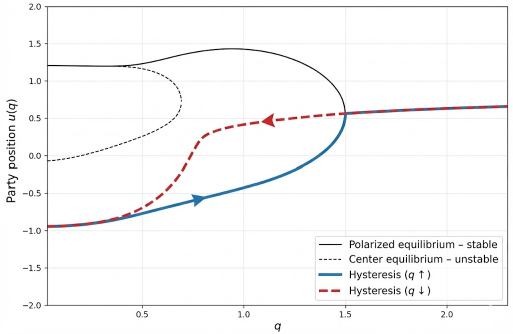}
      \caption{Hysteresis in party position $u(q)$ as tolerance $q$ varies.}
      \label{fig:hysteresis-loop}
    \end{subfigure}

    \caption{Asymmetric electorate and hysteresis. Panel~\subref{fig:voter-gaussian-mixture}
    shows the voter density as a mixture of two Gaussian distributions. Panel~\subref{fig:hysteresis-loop} shows the equilibrium position of
party~1 as a function of $q$ for the asymmetric Gaussian mixture with
$(d_1,\sigma_1) = (1,0.8)$ and $(d_2,\sigma_2) = (-1.2,1.4)$, highlighting a hysteresis loop in $u(q)$. The solid blue curve shows the response for increasing $q$: it follows the lower asymmetric stable branch up to the left fold and then switches to the central stable branch. The dashed red curve shows the response for decreasing $q$, tracking the central branch (stable at first and then unstable) down to the right fold, where it connects back to the asymmetric branch.
}

    \label{fig:gaussian-hysteresis}
  \end{figure}

  In the asymmetric specification with general $(d_{1},d_{2},\sigma_{1},\sigma_{2}
  )$, we no longer have closed--form expressions for $u_{\ast}$ and $q_{c}$,
  but the explicit first--order conditions above allow us to compute the equilibrium
  branches numerically and to visualize how asymmetry in the turnout mixture deforms
  the pitchfork structure and can generate hard polarization with hysteresis; see
  Fig.~\ref{fig:gaussian-hysteresis}.

\begin{figure}[htbp]
    \centering
    \includegraphics[width=0.55\textwidth]{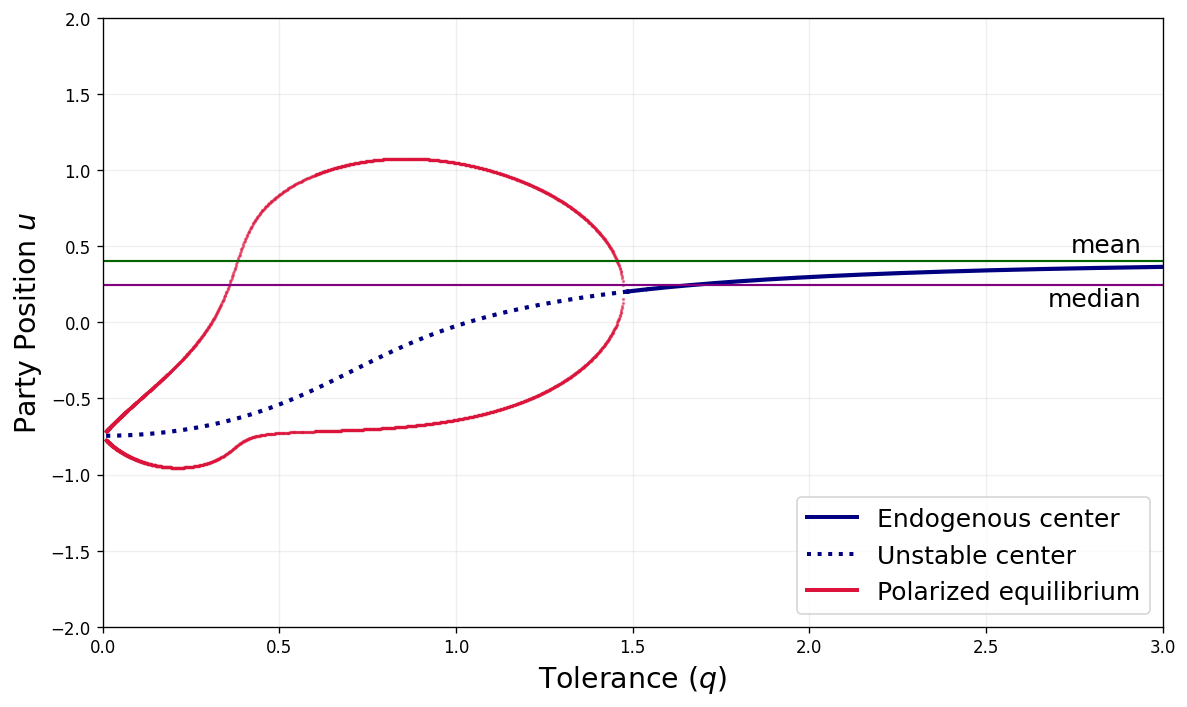}
    \caption{Bifurcation diagram for the satisficing spatial model with
$(d_1,\sigma_1) = (1,0.8)$ and $(d_2,\sigma_2) = (-1.2,1.4)$.
The solid blue curve shows the symmetric equilibrium branch $u_{1} = u_{2}$,
while the solid red curve depicts the asymmetric equilibrium branches.
For $q \ge 1$, the dashed green and dotted magenta horizontal lines indicate,
respectively, the mean and the median of the voter distribution.
}
    \label{fig:bifurcation-mean-median}
  \end{figure}

In the high-tolerance limit $q\rightarrow\infty$, the first-order conditions 
$\partial_u V_{1,q}(u,v)=0$ and $\partial_v V_{2,q}(v,u)=0$ reduce to the linear system  
\begin{align*}
\sigma_{1}(u+d_{1})+\sigma_{2}(u+d_{2}) &= 0, \\
\sigma_{1}(v+d_{1})+\sigma_{2}(v+d_{2}) &= 0,
\end{align*}
whose solution is the mean value $\overline{x}$ of the voter density $\rho(x)$, as predicted by Lemma \ref{lem:large-q-limit}:
\[
u = v = \bar{x}
= -\frac{d_{1}\sigma_{1} + d_{2}\sigma_{2}}{\sigma_{1} + \sigma_{2}}
\quad \text{as } q \to \infty .
\]
Fig.~\ref{fig:bifurcation-mean-median} illustrates this result: as $q$ increases, both equilibrium solutions collapse onto $u = v$, asymptotically approaching the mean value $\bar{x}$ rather than the median—as would be expected in classical spatial models. In the case where $\rho(x)$ is symmetric, $\sigma_{1} = \sigma_{2}$, and $d_{1} = -d_{2}$, the mean $\bar{x} = 0$ coincides with the median, which is the parties’ equilibrium position for $q > q_{c}$.

\section{Proof of the Symmetric Case}\label{sec:symmetric}

We now focus on the symmetric case characterized by an even voter density $\rho$. The resulting one-dimensional reduction yields explicit integral formulas for the bifurcation coefficients, allowing for a rigorous analysis of the linear and cubic terms as the tolerance parameter $q$ varies. We prove that this system undergoes a single symmetric pitchfork bifurcation at a critical value $q_c$ and that above this threshold, the unique stable equilibrium is the distribution's median, which in this symmetric context coincides with the mean.

By Lemma~\ref{lem:parity}, $F_q^{+}(m,0)$ is an odd function
of $m$, so $m=0$ is a zero for all $q>0$. Under our standing assumption that this zero is simple and unique, the central equilibrium branch is given by $m_0(q)\equiv 0$, and the two–dimensional system $(F_q^{+},F_q^{-})=(0,0)$
reduces to a single scalar equation in the polarization distance $d$. Since $F_q^{+}$ is even in $d$, we have $F_q^{+}(0,d)\equiv 0$ for all $d$, so the bifurcation dynamics are entirely governed by
\[
  F_q^{-}(0,d)=0,
\]
which is a one–dimensional problem in the distance parameter $d$.

Let the satisficing function be $f_q(x):=f(x/q)$ and let the voter density $\rho$ be even. Since the satisficing kernel is identical for both parties, we write the payoff of party~1 simply as $V_q(u,v)$, whose expected vote share we know to be given by
\[
  \label{eq:Vq-def}
  V_q(u,v)
  = \int_{\mathbb{R}}\Bigl[f_q(|x-u|)
   - \tfrac{1}{2}f_q(|x-u|)f_q(|x-v|)\Bigr]\rho(x)\,dx .
\]
We then consider the derivative of this payoff along the anti-diagonal solution $v=-u$
and set
\begin{align*}
  \Psi_q(u)
  &:= \frac{\partial V_q}{\partial u}(u,-u)
  \\
  &= -\frac{1}{q}\int_{\mathbb{R}}
    f'\Bigl(\tfrac{|x-u|}{q}\Bigr)\,\operatorname{sgn}(x-u)\,
    \Bigl(1-\tfrac12 f\!\left(\tfrac{|x+u|}{q}\right)\Bigr)\rho(x)\,dx .
\end{align*}

Using the symmetry $V_{2,q}(u,v)=V_q(v,u)$ and the evenness of $\rho$, and observing that along the branch $m=0$ one has $d=u$, a change of variables shows that $\Psi_q$ is odd and that
\[
  F_q^{-}(0,u)
  = \tfrac12\bigl(\Psi_q(u)-\Psi_q(-u)\bigr)
  = \Psi_q(u)
  \qquad\text{for all }u\in\mathbb{R}.
\]
Hence, in the symmetric benchmark, the bifurcation analysis reduces to studying the zeros of the odd function $\Psi_q$, for which $u=0$ is a solution for all $q>0$.

  \subsection{The Pitchfork Bifurcation Condition}

  A pitchfork bifurcation occurs at $u=0$ if this trivial solution changes stability. Since $\Psi_{q}(u)$ is odd, its Taylor expansion around $u=0$ is
  \[
    \Psi_{q}(u)=\mu(q)\,u+\beta(q)\,u^{3}+o(u^{3}), \qquad u\to0.
  \]
  Note that in this symmetric context where $m_{0}(q)\equiv 0$ and the branch curvature vanishes, $\beta_{\mathrm{eff}}(q)$ coincides with $\beta(q)$.

  \begin{theorem}[1D Pitchfork]
    \label{thm:1D-pitchfork} Let $\Psi_{q}(u)$ be $C^{3}$ with $\Psi_{q}(0)=0$ for
    all $q$. A pitchfork bifurcation occurs at $(u,q)=(0,q_{c})$ if
    $q_{c}$ satisfies:
    \begin{enumerate}
      \item[(a)] \textit{Linear Transversality:} The linear coefficient
        $\mu(q) := \Psi_{q}'(0)$ crosses zero transversally at $q=q_c$:
        \[
          \mu(q_{c})=0 \quad \text{and}\quad \mu'(q_{c}) \neq 0.
        \]

      \item[(b)] \textit{Cubic Non-degeneracy:} The cubic coefficient
        $\beta(q) := \frac{1}{6}\Psi_{q}'''(0)$ is non-zero at $q_{c}$:
        \[
          \beta(q_{c}) \neq 0.
        \]
    \end{enumerate}
    The signs of $\mu'(q_{c})$ and $\beta(q_{c})$ determine if the bifurcation is supercritical or subcritical.
  \end{theorem}

  Our task is to verify these two conditions for our model.

  \subsection{Calculation of the Linear Coefficient \texorpdfstring{$\mu(q)$}{mu(q)}}

  \begin{proposition}
    The linear coefficient $\mu(q) = \Psi_{q}'(0)$ is given by
    \begin{equation}
      \; \mu(q)=\frac{2}{q}\int_{0}^{\infty}\Bigl[ f''(t)\Bigl(1-\tfrac12 f(t)\Bigr)+\tfrac12\bigl(f'(t)\bigr)^{2}\Bigr]\, \rho(qt)\,dt. \;
      \label{muqsymm}
    \end{equation}
    In the symmetric benchmark where $m_{0}(q)\equiv 0$, this definition coincides with the general definition
    $\mu(q)=\partial_{d} F_{q}^{-}(m,0)\big|_{m=0}$.
  \end{proposition}

  \begin{proof}
    We differentiate $\Psi_{q}(u)$ at $u=0$ using the product rule inside the integral.
    Let
    \[
      \Psi_{q}(u)\;=\;-\!\int_{\mathbb{R}}\underbrace{\tfrac1q\,f'\,\!\Big(\tfrac{|x-u|}{q}\Big)\operatorname{sgn}(x-u)}
      _{=:A(u,x)}\, \underbrace{\Big(1-\tfrac12 f\!\Big(\tfrac{|x+u|}{q}\Big)\Big)}
      _{=:B(u,x)}\, \rho(x)\,dx .
    \]
    Then
\[
\mu(q)
= \Psi_{q}'(0)=-\int_{\mathbb{R}}
\left.\partial_{u}\,\!\big[A(u,x)B(u,x)\big]\right|_{u=0}
\,\rho(x)\,dx.
\]

    \textit{Term (I): $\partial_{u} A(u,x)|_{u=0}\, B(0,x)$.} Using
    distributional derivatives:
    \[
      \frac{\partial}{\partial u}\Big(f'\,\!\Big(\tfrac{|x-u|}{q}\Big)\operatorname{sgn}
      (x-u)\Big) = -\frac{1}{q}\,f''\,\!\Big(\tfrac{|x-u|}{q}\Big)-2\,f'\,\!\Big(\tfrac
      {|x-u|}{q}\Big)\delta(x-u).
    \]
    Multiplying by $1/q$ and evaluating at $u=0$:
    \begin{align*}
      \partial_{u} A(u,x)|_{u=0}&= \Big[\partial_{u}\,\!\Big(\tfrac1q f'\Big(\tfrac{|x-u|}
      {q}\Big) \operatorname{sgn}(x-u)\Big)\Big]_{u=0}\\
      &= -\tfrac{1}{q^2}\,f''\,\!\Big(\tfrac
      {|x|}{q}\Big)-\tfrac{2}{q}\,f'\,\!\Big(\tfrac{|x|}{q}\Big)\delta(x).
    \end{align*}
    The integral for Term (I) is
    \[
      -\!\int\Big[-\tfrac{1}{q^2}f''\,\!\Big(\tfrac{|x|}{q}\Big)-\tfrac{2}{q}\,f'\,\!
      \Big(\tfrac{|x|}{q}\Big)\delta(x)\Big] \Big(1-\tfrac12 f\,\!\Big(\tfrac{|x|}{q}
      \Big)\Big)\rho(x)\,dx.
    \]
    The term with $\delta(x)$ evaluates at $x=0$, but $f'(0)=0$ (since $f$ has a
    maximum at 0), so this term vanishes. We are left with
    \[
      \text{(I)}: \frac{1}{q^{2}}\int_{\mathbb{R}}f''\,\!\Big(\tfrac{|x|}{q}\Big)\Big
      (1-\tfrac12 f\,\!\Big(\tfrac{|x|}{q}\Big)\Big)\rho(x)\,dx.
    \]

    \textit{Term (II): $A(0,x) \partial_{u}B(u,x)|_{u=0}.$} Differentiating $B(u,x)$ with respect to $u$ and evaluating at $u=0$, we obtain
\begin{align*}
      \partial_{u}B(u,x)|_{u=0}&= \Big[-\tfrac{1}{2q}\,f'\,\!\Big(\tfrac{|x+u|}{q}\Big)\operatorname{sgn}
    (x+u)\Big]_{u=0}\\
      &= -\tfrac{1}{2q}\,f'\,\!\Big(\tfrac{|x|}{q}\Big)\operatorname{sgn}
      (x).
    \end{align*}
    Also, $A(0,x) = \tfrac1q f'\,\!\Big(\tfrac{|x|}{q}\Big)\operatorname{sgn}(x)$.
    The integral for Term (II) is
    \[
      -\!\int\!\Big[\tfrac1q f'\,\!\Big(\tfrac{|x|}{q}\Big)\operatorname{sgn}(x)\Big
      ]\, \Big[-\tfrac{1}{2q}\,f'\,\!\Big(\tfrac{|x|}{q}\Big)\operatorname{sgn}(x)\Big
      ]\,\rho(x)\,dx.
    \]
    Since $\operatorname{sgn}(x)^{2}=1$ (for $x\neq 0$, which is sufficient for integration),
    this simplifies to
    \[
      \text{(II)}: \tfrac{1}{2q^2}\int_{\mathbb{R}}\Big(f'\,\!\Big(\tfrac{|x|}{q}\Big
      )\Big)^{2}\rho(x)\,dx.
    \]

    Combining Terms (I) and (II) yields the expression for $\mu(q)$:
\[
      \mu(q) =\frac{1}{q^{2}}\int_{\mathbb{R}}\bigg[ f''\,\!\Big(\tfrac{|x|}{q}\Big
      )\Big(1-\tfrac12 f\,\!\Big(\tfrac{|x|}{q}\Big)\Big) +\tfrac12\Big(f'\,\!\Big(\tfrac
      {|x|}{q}\Big)\Big)^{2}\bigg]\,\rho(x)\,dx.
    \]
    Letting $t=x/q$ and using the parity of $\rho$, this reduces to Eq. (\ref{muqsymm}).
  \end{proof}

  \subsection{Existence of the Bifurcation Point \texorpdfstring{$q_{c}$}{q*}}

  We now show that $\mu(q)$ must cross zero, guaranteeing the existence of $q_{c}$.

  \begin{theorem}[Sign Change of $\mu(q)$]
    Let $f$ be $C^{2}$ with $f(0)=1, f'(0)=0, f''(0)<0$,
    $\lim_{t\to\infty}f(t)=0$, and $f' \in L^{2}(0,\infty)$. Let $\rho$ be an
    even, bounded probability density, continuous at the origin with $\rho(0)>0$.
    Define
    \[
      \psi(t)\;:=\;f''(t)\!\left(1-\tfrac12 f(t)\right)+\tfrac12\big(f'(t)\big)^{2}
      .
    \]
    Then the linear coefficient $\mu(q)\;:=\;\frac{2}{q}\int_{0}^{\infty}\psi(t )
    \,\rho(qt)\,dt$ is continuous on $(0,\infty)$ and satisfies
    \[
      \lim_{q\to 0^+}q \mu(q) = 2\rho(0)\int_{0}^{\infty}(f'(t))^{2}dt > 0 \quad
      (\,\mu(q) > 0 \text{ as }q \to 0^{+}),
    \]
    \[
      \lim_{q\to \infty}q^{2}\mu(q) = \psi(0) = \tfrac{1}{2}f''(0) < 0 \qquad (\,\mu(q) < 0 \text{ as }q \to \infty).
    \]
    By the Intermediate Value Theorem, there must exist at least one
    $q_{c}>0$ such that $\mu(q_{c})=0$.
    \label{signchangesymm}
  \end{theorem}

\begin{remark}[Supercritical Bifurcation]
The sign change of $\mu(q)$ established in Theorem~\ref{signchangesymm} implies that, if $q_c$ is a simple zero of $\mu(q)$---and in particular if it is the first positive root---then there exists a neighborhood of $q_c$ such that $\mu(q)>0$ for $q<q_c$ and $\mu(q)<0$ for $q>q_c$. In this case one necessarily has $\mu'(q_c)<0$.
Consequently, under the nondegeneracy condition $\beta(q_c)\neq 0$, the additional requirement $\beta(q_c)<0$ guarantees that the pitchfork bifurcation at $q=q_c$ is
supercritical. In particular, there exists a neighborhood of $q_c$ in which the central equilibrium branch is locally stable for $q>q_c$, whereas for $q<q_c$ it loses stability and two symmetric polarized equilibrium branches emerge, both locally stable.
\end{remark}

  \begin{proof}
    \textit{Structural Identity.} Let $U(t):=f'(t)(1-\tfrac12 f(t))$. Then $U'(t)
    =f''(t)(1-\tfrac12 f(t))-\tfrac12(f'(t))^{2}$. Thus, $\psi(t)=U'(t)+(f'(t))^{2}$.
    Integrating and using boundary conditions ($f'(0)=0$, $f'(\infty)=0$, $f(\infty
    )=0$), we find
    \[
      \int_{0}^{\infty}\psi(t)\,dt =\big[U(t)\big]_{0}^{\infty}+\int_{0}^{\infty}
      \big(f'(t)\big)^{2}\,dt =\int_{0}^{\infty}\big(f'(t)\big)^{2}\,dt.
    \]
    Since $f$ is not constant, this integral is strictly positive. We also assume
    $f' \in L^{2}$, so the integral is finite.

    \textit{Continuity of $\mu(q)$.} Let $I(q) := \int_{0}^{\infty}\psi(s/q) g (s
    ) ds$. Since $\psi$ is bounded (as $f, f', f''$ are) and $\rho \in L^{1}$, the
    Dominated Convergence Theorem implies $I(q)$ is continuous for $q>0$.
    Since $\mu(q) = (2/q^{2})I(q)$, $\mu(q)$ is also continuous for $q>0$.

    \textit{Asymptotic for $q\to 0^{+}$.} We analyze
    $q \mu(q) = 2 \int_{0}^{\infty}\psi(t) \rho(qt) dt$. As $q\to 0^{+}$,
    $\rho(qt) \to \rho(0)$ for each $t$. A direct application of the Dominated Convergence
    Theorem is not immediate, so a more careful analysis is needed. Let $T(q) = q
    ^{-1/2}$, so $T(q) \to \infty$ and $qT(q) \to 0$. We split the integral:
    \[
      \frac{q \mu(q)}{2}= \int_{0}^{T(q)}\psi(t) \rho(qt) dt + \int_{T(q)}^{\infty}
      \psi(t) \rho(qt) dt.
    \]
    Let $\omega(r):=\sup_{|x|\le r}\,|\rho(x)-\rho(0)| \to 0$ as $r\to 0$. The first
    term is $\int_{0}^{T(q)}\psi(t) (\rho(0) + (\rho(qt)-\rho(0))) dt$. The
    error is bounded by
    \begin{align*}
      \Bigl|\int_{0}^{T(q)}\psi(t) (\rho(qt)-\rho(0)) dt\Bigr| &\le \omega(qT(q))
      \int_{0}^{T(q)}|\psi(t)| dt\\
      &\le \omega(q^{1/2}) \|\psi\|_{1}\to 0.
    \end{align*}

    The second term is bounded by
    \[
      \|\rho\|_{\infty}\int_{T(q)}^{\infty}|\psi(t)| dt \to 0
    \]
    since $T(q)\to\infty$ and $\psi \in L^{1}$. Therefore,
    \begin{align*}
      \lim_{q\to 0^+}q \mu(q)& = 2 \lim_{q\to 0^+}\int_{0}^{T(q)}\psi(t) \rho(0) d
      t \\
      &= 2\rho(0) \int_{0}^{\infty}\psi(t) dt = 2\rho(0) \int_{0}^{\infty}(f'(t)
      )^{2}dt > 0.
\end{align*}

    \textit{Asymptotic for $q\to\infty$.} We use the form
    $\mu(q) = \frac{2}{q^{2}}\int_{0}^{\infty}\psi(s/q)\rho(s)\,ds$. As $q\to\infty$,
    $\psi(s/q) \to \psi(0)$ for each $s$. The integrand is dominated by $\|\psi\|
    _{\infty}|\rho(s)|$, which is in $L^{1}(0,\infty)$ since $\rho \in L^{1}$. By dominated convergence:
    \[
      \lim_{q\to\infty}q^{2}\mu(q) = 2 \int_{0}^{\infty}\lim_{q\to\infty}\psi (s/
      q) \rho(s) ds = 2 \int_{0}^{\infty}\psi(0) \rho(s) ds.
    \]
    Using
    $\psi(0) = f''(0)(1-\tfrac{1}{2}f(0)) + \tfrac{1}{2}(f'(0))^{2}= \tfrac{1}{2}
    f''(0)$, and $\int_{0}^{\infty}\rho(s) ds = \tfrac{1}{2}$ (as $\rho$ is an
    even probability density):
    \[
      \lim_{q\to\infty}q^{2}\mu(q) = 2 \cdot (\tfrac{1}{2}f''(0)) \cdot (\tfrac{1}
      {2}) = \tfrac{1}{2}f''(0) < 0.
    \]

    Since $\mu(q)$ is continuous, positive for small $q$, and negative for large
    $q$, the Intermediate Value Theorem yields at least one $q_{c}>0$ such
    that $\mu(q_{c})=0$.
  \end{proof}

  \begin{remark}[Transversality Condition]
    The second condition for the bifurcation is that $\mu'(q_{c}) \neq 0$. Let
    $D(q) := q\mu(q)/2$. Differentiating yields
    $\mu'(q_{c}) = 2D'(q_{c})/q_{c}$. Thus,
    $\mu'(q_{c}) \neq 0$ is equivalent to $D'(q_{c}) \neq 0$:
    \begin{equation}
      D'(q_{c}) = \frac{d}{dq}\int_{0}^{\infty}\psi(t) \rho(qt) dt \Big|_{q=q_{c}}
      = \int_{0}^{\infty}\psi(t) t \rho'(q_{c}t) dt \neq 0.
      \label{Ddiff}
    \end{equation}
    This is a generic nondegeneracy condition on the derivatives of  the density  $\rho$, and the validity of the differentiation under the integral sign can be ensured under the additional regularity assumption that $\rho$ is absolutely continuous on
$\mathbb{R}$ and satisfies $\int_{0}^{\infty} t\,|\rho'(t)|\,dt<\infty$. To see this, fix $q>0$ and consider the difference quotient:
\[
\frac{D(q+h)-D(q)}{h}
=\int_{0}^{\infty}\psi(t)\,\frac{\rho((q+h)t)-\rho(qt)}{h}\,dt.
\]
Since $\rho$ is absolutely continuous, for a.e. $t\ge 0$ we have
\[
\frac{\rho((q+h)t)-\rho(qt)}{h}
=t\int_{0}^{1}\rho'\bigl((q+\theta h)t\bigr)\,d\theta.
\]
Therefore,
\[
\left|\psi(t)\,\frac{\rho((q+h)t)-\rho(qt)}{h}\right|
\le \|\psi\|_{\infty}\, t\int_{0}^{1}\left|\rho'\bigl((q+\theta h)t\bigr)\right|\,d\theta.
\]
For $|h|$ sufficiently small we have $q+\theta h\ge q/2$ for all $\theta\in[0,1]$.
Hence, for each $\theta\in[0,1]$ and such $h$, using the change of variables
$\tau=(q+\theta h)t$ we obtain
\[
\begin{aligned}
\int_{0}^{\infty} t\,\left|\rho'\bigl((q+\theta h)t\bigr)\right|\,dt
&= \frac{1}{(q+\theta h)^2}\int_{0}^{\infty}\tau\,|\rho'(\tau)|\,d\tau \\
&\le \frac{4}{q^{2}}\int_{0}^{\infty}\tau\,|\rho'(\tau)|\,d\tau < \infty .
\end{aligned}
\]
Consequently, the integrands in the difference quotient are dominated by an $L^{1}(0,\infty)$ function (uniformly in $\theta$ and for small $h$), and we may apply the Dominated Convergence Theorem to pass to the limit $h\to 0$ inside the integral. Since
\[
\lim_{h\to 0}\frac{\rho((q+h)t)-\rho(qt)}{h}=t\,\rho'(qt)\qquad\text{for a.e.\ }t,
\]
we conclude that $D$ is differentiable and that Eq. (\ref{Ddiff}) holds.
  \end{remark}

\subsection{Calculation of the Cubic Coefficient \texorpdfstring{$\beta(q)$}{mu(q)}}

  \begin{proposition}
The cubic coefficient $\beta(q) = \frac{1}{6}\Psi_q'''(0)$ is given by
\begin{equation}
\beta(q)
  =\frac{f^{(3)}(0)\rho(0)}{6q^{3}}
   -\frac{2}{q^{3}}\int_{0}^{\infty}\eta(t)\rho(qt)\,dt,
   \label{betaprop}
\end{equation}
where $\eta(t)$ is defined by
\begin{equation}
\eta(t):=\frac{1}{4}\bigl(f''(t)\bigr)^{2}
-\frac{1}{3}f'(t)f^{(3)}(t)
-\frac{1}{6}f^{(4)}(t)\Bigl(1-\tfrac12 f(t)\Bigr).
\label{etaprop}
\end{equation}
   \end{proposition}

\begin{proof}
Let
    \[
      \Psi_{q}(u)\;=\;-\!\int_{\mathbb{R}}\underbrace{\tfrac1q\,f'\,\!\Big(\tfrac{|x-u|}{q}\Big)\operatorname{sgn}(x-u)}
      _{=:A(u,x)}\, \underbrace{\Big(1-\tfrac12 f\,\!\Big(\tfrac{|x+u|}{q}\Big)\Big)}
      _{=:B(u,x)}\,\rho(x)dx
    \]
and consider the Taylor expansions $A(u,x)=\sum_{k=0}^{3}A_{k}(x)u^k+O(u^4)$ and $B(u,x)=\sum_{k=0}^{3}B_{k}(x)u^k+O(u^4)$ around $u=0$. Using distributional derivatives to handle the absolute values (with no boundary contribution thanks to $f'(0)=0$) and collecting the $u^3$ terms in the product $A(u,x)B(u,x)$, we obtain
\[
\begin{aligned}
&A_{3}\,B_{0} \;=\; -\frac{1}{6q^4}\,\Bigl[f^{(4)}\!\left(\tfrac{|x|}{q}\right)+2qf^{(3)}\!\left(\tfrac{|x|}{q}\right)\delta(x)\Bigr]\,\Bigl(1-\tfrac12 f\!\left(\tfrac{|x|}{q}\right)\Bigr),\\
&A_{2}\,B_{1} \;=\; -\,\frac{1}{4q^4}\,f^{(3)}\!\left(\tfrac{|x|}{q}\right)\,f'\!\left(\tfrac{|x|}{q}\right),\\
&A_{1}\,B_{2} \;=\; +\,\frac{1}{4q^4}\,\Bigl(f''\!\left(\tfrac{|x|}{q}\right)\Bigr)^2,\\
&A_{0}\,B_{3} \;=\; -\,\frac{1}{12q^4}\,f'\!\left(\tfrac{|x|}{q}\right)\,f^{(3)}\!\left(\tfrac{|x|}{q}\right).
\end{aligned}
\]
Grouping all these terms, the cubic contribution in $u$ to $A(u,x)B(u,x)$ is
\[
\frac{u^3}{q^4}\left\{\eta\left(\frac{|x|}{q}\right)-\frac{q}{3}f^{(3)}\left(\frac{|x|}{q}\right)\delta(x)\left[1-\frac{1}{2}f\left(\frac{|x|}{q}\right)\right]\right\}.
\]
Since $\Psi_{q}(u)=-\!\int_{\mathbb{R}} A(u,x)B(u,x)\,\rho(x)dx$, the cubic coefficient $\beta(q) = \frac{1}{6}\Psi_q'''(0)$ is given by Eqs. (\ref{betaprop}) and (\ref{etaprop}), upon using the parity of $\rho(x)$ and the change of variable $x=qt$.
\end{proof}

  \subsection{Summary of Bifurcation Conditions}
  A pitchfork bifurcation occurs at $q=q_{c}$ if $\mu(q_{c})=0$,
  $\mu'(q_{c}) \neq 0$, and $\beta(q_{c}) \neq 0$. In terms of the
  integral conditions, this translates to:
  \begin{enumerate}
    \item $\mu(q_{c})=0$:
      \[
        \int_{0}^{\infty}\underbrace{\left[ f''(t)\Bigl(1-\tfrac12 f(t)\Bigr) + \tfrac12
        \bigl(f'(t)\bigr)^2 \right]}_{=:\psi(t)}\,\rho(q_{c}t)\,dt = 0
      \]

    \item $\mu'(q_{c})\neq 0$:
      \[
        \int_{0}^{\infty}\psi(t) t \rho'(q_{c}t)\,dt \neq 0
      \]

    \item $\beta(q_{c})\neq 0$:
    {\small 
      \[
        \int_{0}^{\infty}\Biggl[ \frac{1}{4}\,\bigl(f''(t)\bigr)^{2}-\frac{1}{3}\,
        f'(t)\,f^{(3)}(t) -\frac{1}{6}\,f^{(4)}(t)\Bigl(1-\tfrac12 f(t)\Bigr) \Biggr
        ]\,\rho(q_{c}t)\,dt\]
        \[\neq \frac{f^{(3)}(0)\rho(0)}{12}
      \] }%

      \item If $q_c$ is a simple zero of $\mu(q)$—and in particular if it is the first positive root—then $\mu'(q_c)<0$, and the additional requirement
      \[ \beta(q_c)<0 \]ensures that the pitchfork bifurcation at $q=q_c$ is \textit{supercritical}. Consequently, the central equilibrium is locally stable for $q>q_c$, while two symmetric polarized equilibria are locally stable for $q<q_c$.

  \end{enumerate}

  \section{Symmetric Electorate Examples}
  \label{sec:symexamples}

In this section we specialize the abstract bifurcation theory developed above to three concrete choices of satisficing kernel $f$ and voter density $\rho$ in the symmetric case $\rho(x)=\rho(-x)$. For each pair $(f,\rho)$ we compute the linear and cubic coefficients $\mu(q)$ and $\beta(q)$, identify a critical value $q_{c}>0$ such that $\mu(q_{c})=0$, and verify the non-degeneracy conditions $\mu'(q_{c}) \neq 0$ and $\beta(q_{c}) \neq 0$, which guarantee the occurrence of a pitchfork bifurcation in the reduced one–dimensional equation for the polarization distance. Throughout this section we use the general formula \eqref{muqsymm} for $\mu(q)$ and the expressions \eqref{betaprop}–\eqref{etaprop} for $\beta(q)$. In particular, in all examples considered below one has $\mu'(q_c)<0$ and $\beta(q_c)<0$, so that the pitchfork bifurcation is supercritical.

  \subsection{Case 1: Gaussian--Gaussian} 

We first consider the case in which both the satisficing kernel and the voter density are Gaussian. For this choice, straightforward calculation using Eq. (\ref{muqsymm}) yields the explicit expression for the linear coefficient
\[
  \mu(q)=\frac{1}{2q\sqrt{q^{2}+2}}-\frac{q}{\left(q^{2}+1\right)^{3/2}}.
\]
The critical value of the tolerance parameter $q$ is therefore obtained as the root of $\mu(q)$, which is equivalent to solving (e.g., \cite{YAKM})
\[
  (q_{c}^{2}+1)^{3}-4q_{c}^{4}(q_{c}^{2}+2)=0.
\]
The bifurcation parameter is thus given by the unique positive root of this equation,
\[
  q_{c}\approx 0.807379,
\]
for which $\mu'(q_c)<0$.

The cubic coefficient is given by Eqs. (\ref{betaprop}) and (\ref{etaprop})
\[
  \beta(q)=\frac{q}{2 \left(q^{2}+1\right)^{5/2}}-\frac{1}{2 q^{3} \sqrt{q^2+2}},
\]
and satisfies $\beta(q_c)<0$, so that the pitchfork bifurcation is supercritical, with two stable symmetric polarized equilibria emerging for $q<q_c$.

  \subsection{Case 2: Cauchy--Gaussian} 

  We now turn to a mixed specification in which the satisficing kernel is Cauchy-shaped while the voter density remains Gaussian. For this pair, the linear coefficient admits the closed form given by Eq. (\ref{muqsymm})
  {\small
  \[
    \mu(q)=\frac{e^{\frac{q^2}{2}}\left(q^{6}+24 q^{4}+27 q^{2}-6\right) \sqrt{\frac{\pi}{2}}\Bigl[\operatorname{erf}\!\left(\frac{q}{\sqrt{2}}\right)-1\Bigr]
    +\Bigl(q^{5}+23q^{3}+6q\Bigr)}{24q}.
  \]
  }%
  The smallest positive root of $\mu(q)$ determines the bifurcation parameter
  \[
    q_{c}\approx 0.975328,
  \]
  for which $\mu'(q_c)<0$.

The cubic coefficient follows from Eqs. (\ref{betaprop}) and (\ref{etaprop})
\[
\beta(q)= -\frac{1}{240 q^{3}}
\left[
e^{q^{2}/2} P_{1}(q)
\sqrt{\pi/2}\,\bigl(\operatorname{erf}(q/\sqrt{2})-1\bigr)+ q P_{2}(q)
\right],
\]
where $P_1(q)$ and $P_2(q)$ are given by
\begin{equation}
\begin{aligned}
P_{1}(q) &= q^{10}+45q^{8}+245q^{6}+105q^{4}+60q^{2}-120,
\\
P_{2}(q) &= q^{8}+44q^{6}+203q^{4}-20q^{2}+120,
\end{aligned}\nonumber
\end{equation}
and satisfies $\beta(q_c)<0$, implying that the pitchfork bifurcation is supercritical, giving rise to polarization as tolerance decreases below $q_c$.

  \subsection{Case 3: Cauchy--Cauchy} 

  \begin{figure}[h]
    \centering

    \begin{subfigure}
      {0.49\textwidth}
      \centering
      \includegraphics[width=\linewidth]{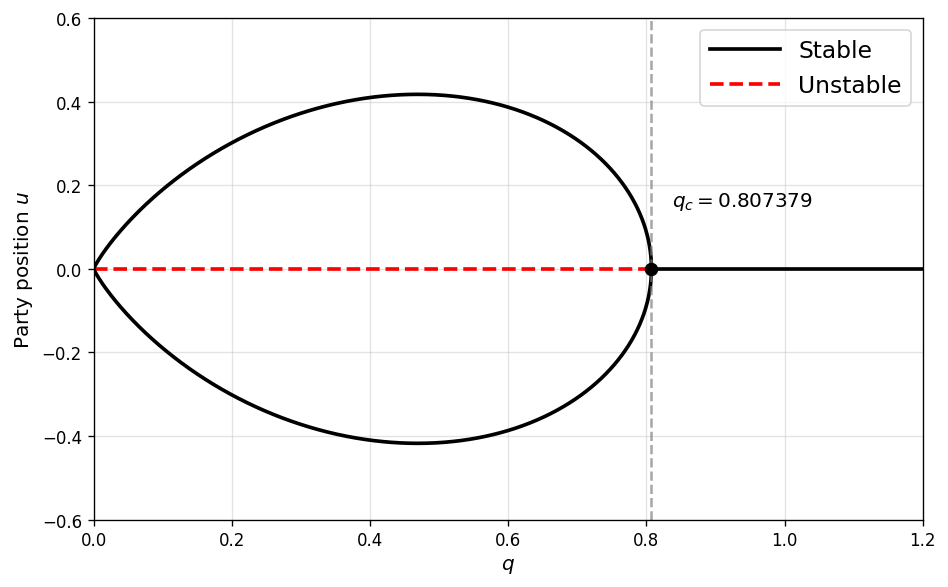} 
      \caption{Gaussian–Gaussian case.}
      \label{fig:bifurc-gg}
    \end{subfigure}
    \hfill
    \begin{subfigure}
      {0.49\textwidth}
      \centering
      \includegraphics[width=\linewidth]{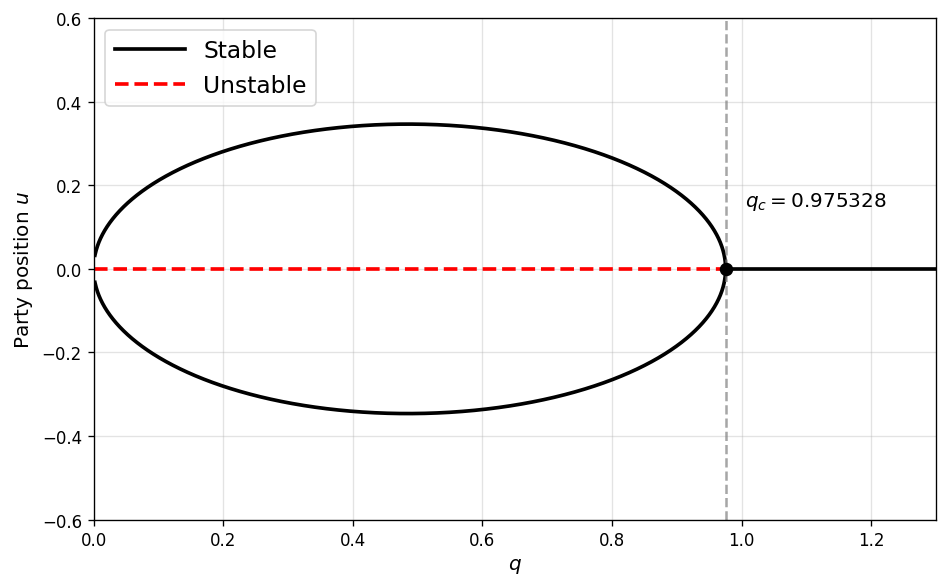} 
      \caption{Cauchy kernel, Gaussian electorate.}
      \label{fig:bifurc-cg}
    \end{subfigure}
    \caption{\textit{Bifurcation diagrams for representative kernel–electorate pairs.} Each panel shows the symmetric equilibrium branch $u(q)$, defined by the implicit equilibrium condition $\Psi_q(u)=0$, as a function of the tolerance parameter $q$, with the analytically derived critical value $q_c$ indicated by a dashed vertical line. All cases display a supercritical pitchfork bifurcation.}
    \label{fig:bifurcation-comparison}
  \end{figure}

  Finally, we consider the Cauchy-Cauchy specification, in which both the satisficing kernel and the voter density are heavy-tailed. In this case, the linear coefficient reduces to the rational expression given by Eq.~(\ref{muqsymm}).
  \[
    \mu(q)=\frac{1-2q(2q^{2}+q-2)}{4q(q+1)^{4}}.
  \]
  The critical value $q_{c}$ is determined by the cubic equation in the numerator, and its only positive root is given numerically by
  \[
    q_{c}\approx 0.906803,
  \]
for which $\mu'(q_c)<0$.

The cubic coefficient likewise follows from Eqs.~(\ref{betaprop}) and (\ref{etaprop}), also yielding a rational expression.
  \[
    \beta(q)=\frac{2 q^{5}-7 q^{4}-20 q^{3}-15 q^{2}-6 q-1}{2q^{3}(q+1)^{6}},
  \]
and satisfies $\beta(q_c)<0$, so that the pitchfork bifurcation is supercritical, with the onset of polarization for $q<q_c$.

  \medskip

Fig. \ref{fig:bifurcation-comparison} illustrates the bifurcation diagrams defined by the implicit equilibrium condition $\Psi_q(u)=0$ for the first two kernel–electorate pairs considered in this section, which are representative of the general behavior. In both cases, the analytically predicted critical value $q_c$ marks a supercritical pitchfork bifurcation, with loss of stability of the central equilibrium and the emergence of two stable symmetric polarized branches as $q$ decreases below $q_c$. The third example exhibits the same qualitative bifurcation structure and is therefore omitted for brevity.

\section{Conclusion}
\label{sec:conclusion}

We have investigated a general satisficing spatial model of electoral competition in which ideological divergence arises endogenously from the interaction between voter tolerance and strategic positioning. Our results show that convergence to the median voter is not a universal principle but a regime-specific outcome: in the symmetric benchmark, convergence to the center is lost through a supercritical pitchfork bifurcation once tolerance falls below a critical threshold. This mechanism provides a rigorous account of how ideological separation can emerge without exogenous shocks, even without assuming extremist candidates.

For asymmetric electorates, the same bifurcation can unfold into a hysteresis loop, creating a ``lock-in'' effect in which polarization may persist even after voter tolerance is restored. More broadly, we show that hysteresis can arise in a simple \emph{equation-based} model, complementing computationally intensive \emph{agent-based} approaches to polarization \cite{Macy2021,VA2018,SWJK2024}.

In the high-tolerance limit, parties converge to the mean of the voter distribution rather than the median, suggesting that sufficiently tolerant electorates induce a center-of-mass (mean) logic rather than the classical median-voter logic. Under symmetric voter distributions, however, the mean coincides with the median, so the median voter prediction is recovered for sufficiently large tolerance. These results---established under general regularity conditions---suggest that the stability of the political center depends fundamentally on the geometry of voter satisfaction and on voters' tolerance to ideological distance.

Naturally, the model is deliberately stylized. We assume a one--dimensional ideological space and impose symmetry in party technology by letting both parties share the same satisficing kernel $f$ and tolerance parameter $q$. These choices are made to preserve analytical tractability and to isolate the role of voter tolerance in shaping equilibrium structure. Despite this parsimony, the model is still capable of generating a rich set of qualitative phenomena---including symmetry breaking, asymmetric polarization, and hysteresis---highlighting that complex political dynamics may arise even in minimally structured spatial environments.

Several extensions seem promising for future research. First, extending the bifurcation analysis to multidimensional ideological spaces may uncover richer symmetry-breaking patterns, especially when different dimensions have unequal salience or interact nonlinearly. Second, allowing parties to differ in their satisficing kernels or tolerance parameters could shed light on strategic asymmetries across organizations with distinct activist bases. Third, incorporating voters' perceptual distortions---such as biased or asymmetric representations of ideological distance---may offer a bridge to models of affective polarization. Finally, introducing explicit dynamics in which party positions evolve over time could illuminate how hysteresis interacts with slow social change, shocks, learning, and institutional constraints. Each of these extensions would deepen the connection between the mathematical structure of satisficing behavior and empirical patterns of political polarization.

  \bibliographystyle{elsarticle-num}
  \bibliography{referencias}

@book{GuilleminPollack,
  author    = {Victor Guillemin and Alan Pollack},
  title     = {Differential Topology},
  publisher = {Prentice--Hall},
  address   = {Englewood Cliffs, NJ},
  year      = {1974},
}

@book{Downs1957,
  author    = {Anthony Downs},
  title     = {An Economic Theory of Democracy},
  publisher = {Harper \& Row},
  address   = {New York},
  year      = {1957}
}

@article{Hotelling1929,
  author  = {Harold Hotelling},
  title   = {Stability in Competition},
  journal = {The Economic Journal},
  volume  = {39},
  number  = {153},
  pages   = {41--57},
  year    = {1929}
}

@article{YAKM,
  author  = {Yang, V. C. and Abrams, D. M. and Kernell, G. and Motter, A. E.},
  title   = {Why are {U.S.} parties so polarized? A ``satisficing'' dynamical model},
  journal = {SIAM Review},
  volume  = {62},
  number  = {3},
  pages   = {646--657},
  year    = {2020}
}

@article{Alesina1988,
  author  = {Alberto Alesina},
  title   = {Credibility and Policy Convergence in a Two-Party System with Rational Voters},
  journal = {American Economic Review},
  volume  = {78},
  number  = {4},
  pages   = {796--805},
  year    = {1988}
}

@article{AlesinaCukierman1990,
  author  = {Alberto Alesina and Alex Cukierman},
  title   = {The Politics of Ambiguity},
  journal = {Quarterly Journal of Economics},
  volume  = {105},
  number  = {4},
  pages   = {829--850},
  year    = {1990}
}

@book{AlesinaRosenthal1995,
  author    = {Alberto Alesina and Howard Rosenthal},
  title     = {Partisan Politics, Divided Government, and the Economy},
  publisher = {Cambridge University Press},
  address   = {New York},
  year      = {1995}
}

@article{Aldrich1983,
  author  = {John H. Aldrich},
  title   = {A Downsian Spatial Model with Party Activism},
  journal = {American Political Science Review},
  volume  = {77},
  number  = {4},
  pages   = {974--990},
  year    = {1983}
}

@article{Baron1993,
  author  = {David P. Baron},
  title   = {Government Formation and Endogenous Parties},
  journal = {American Political Science Review},
  volume  = {87},
  number  = {1},
  pages   = {34--47},
  year    = {1993}
}

@book{HinichMunger1994,
  author    = {Melvin J. Hinich and Michael C. Munger},
  title     = {Ideology and the Theory of Political Choice},
  publisher = {University of Michigan Press},
  address   = {Ann Arbor},
  year      = {1994}
}

@article{Venegeroles2025,
  author  = {Roberto Venegeroles},
  title   = {Beyond the median voter: A model of how the ideological dimension shapes party polarization},
  journal = {Physical Review E},
  volume  = {111},
  number  = {5},
  pages   = {054316},
  year    = {2025}
}

@article{Macy2021,
  title     = {Polarization and tipping points},
  author    = {Macy, Michael W. and Ma, Manqing and Tabin, Daniel R. and Gao, Jianxi and Szymanski, Boleslaw K.},
  journal   = {Proceedings of the National Academy of Sciences},
  volume    = {118},
  number    = {50},
  pages     = {e2102144118},
  year      = {2021},
}

@article{VA2018,
  title     = {Threshold $q$-voter model},
  author    = {Vieira, Allan R. and Anteneodo, Celia},
  journal   = {Physical Review E},
  volume    = {97},
  number    = {5},
  pages     = {052106},
  year      = {2018},
}

@article{SWJK2024,
  title     = {Toward Understanding of the Social Hysteresis: Insights From Agent-Based Modeling},
  author    = {Sznajd-Weron, Katarzyna and J{\k{e}}drzejewski, Arkadiusz and Kami{\'n}ska, Barbara},
  journal   = {Perspectives on Psychological Science},
  volume    = {19},
  number    = {2},
  pages     = {511--521},
  year      = {2024},
}
\end{document}